# Morphological effects of leading-edge serrations on the acoustic signatures of mixed flow fan analyzed using novel CFD-informed methods

Jinxin Wang[1,2]（王津新）, Kenta Ishibashi[2]（石橋健太）, Teruaki Ikeda[3]（池田旭彰）, Takeo Fujii[3]（藤井武夫）, Toshiyuki Nakata[2, a]（中田敏是）, Hao Liu (劉浩)[1,2,a]

**AFFILIATIONS**

[1]Shanghai Jiao Tong University and Chiba University International Cooperative Research Center (SJTU-CRICRC), 800 Dongchuan Road, Minhang District, Shanghai 200240, People's Republic of China

[2]Graduate School of Engineering, Chiba University, 1-33, Yayoi-cho, Inage-ku, Chiba 263-8522, Japan

[3]TERAL Inc, 230 Moriwake, Miyuki-cho, Fukuyama-shi, Hiroshima 720-0033, Japan

[a] Author to whom correspondence should be addressed: tnakata@chiba-u.jp and hliu@faculty.chiba-u.jp.

Lead Contact: Professor Hao Liu (hliu@faculty.chiba-u.jp)

# ABSTRACT

Leading-edge (LE) noise is a common source of broadband noise for fans that can be suppressed using appended LE serrations. We conduct an integrated study of the morphological effects of interval, length, and inclination angle of owl-inspired LE serrations on the aeroacoustic characteristics of a mixed flow fan using experiments, computational fluid dynamics (CFD), and the Ffowcs Williams–Hawkings (FWH) analogy. A novel method for surface noise strength (SNS) visualization was developed based on the FWH analogy with large-eddy simulations, and a CFD-informed index *SAPG* is proposed to evaluate the severity of flow separation with pressure gradient, which are verified to be effective in examining the acoustic sources and chordwise separation. Acoustic measurements show the robust tradeoff solving capability of the serrations under various morphologies and the SNS visualizations indicate that the separation-induced LE noise is suppressed considerably. One-third octave analyses suggest that extending serration length can lower separation noise more effectively than shrinking the interval over 100–3k Hz. A smaller interval is more desirable while an optimal length exists in association with tonal noise. Moreover, small inclination angles ($\leq 20°$) enable the deceleration of oncoming flows with stagnation relieved, and consequently, further suppress the LE noise, by a flow-buffering effect. Heavy inclination angles ($\geq 40°$) induce an additional tip vortex, causing high-coherence turbulence impingement noise and resulting in a drastic increase in broadband noise at frequencies exceeding 4k Hz. Our study thus clarifies the morphological effects of LE serrations on aeroacoustic signatures of rotary devices while providing useful methods for acoustic analyses.  *(250 words)*

## I. INTRODUCTION

Fan noise has been at the forefront of research conversations in the field owing to its importance in applications ranging from ventilation, cooling, and heating systems to turbofans of large sizes.[1-3] It is known that the sound emitted from the fan in isolation plays a crucial role in noise generation, although the interaction between the fan blades and the outlet vanes (OGVs) contributes the most in multistage configurations.[1, 4] For the noise radiating from fan alone, aside from the tonal noise, the broadband noise tends to be dominant in the case of low-speed fans that comes from the interaction between turbulent flows and blade surface.[4-6] The best-known broadband component is the trailing-edge (TE) noise, which is the scattering of sound when turbulent eddies interact with the TE.[4, 6] By contrast, leading-edge (LE) noise is often considered as the result of the interaction of the LE with oncoming or ingested turbulence,[5-7] which can also cause an increase in tonal noise.[8] Besides, high-frequency broadband noise can radiate from positions such as the tip of the axial fans due to vortex formation where the clearance allows the flows to roll upside driven by the pressure difference between the pressure and suction sides.[9, 10]

The separation-transition can happen due to the adverse pressure gradient (APG) on the blade suction faces where the flows evolve into the coherent vortices from the roll-ups and eventually break down into turbulence, exhibiting the inviscid Kelvin-Helmholtz (KH) instability.[11] Such an instability may also cause serious LE noise due to the impingement of separation-induced turbulence regardless of those from the upstream.[12] For example, LE noise is found to be the dominant noise source at low flow rates up to the design point in axial flow fans.[13] In such cases, the LE noise may be attributed to separation noise rather than turbulence ingestion noise. A recent study was conducted on an airfoil under near-stall conditions and determined the separation noise range of 100–1k Hz at Reynolds number (Re) of $4 \times 10^5$,[14] which is further confirmed for the case of a mixed flow fan over a range of 40–4k Hz (Re $\approx 4.6 \times 10^5$).[12]

The strategy of passive flow control has been found suitable for suppressing the flow separation, and is often observed in animal movements that achieve sophisticated aerodynamic and acoustic performance.[15, 16] Barn owls have been investigated for decades because of their talent for silent flight facilitated partially by the LE serrations in their 10th primary feathers,[17, 18] which have been extensively investigated and applied to improve the acoustic performance of rotary blades since the 1970s.[19, 20] Moreover, these serrations have complex three-dimensional (3D) features[18, 21] that have inspired several airfoil designs with different morphological characteristics.[22, 23] To this end, by studying the 2D LE-serrated airfoils, researchers have found that the amplitude of the serrations are more influential than the interval (or wavelength for wavy serrations) in reducing the sound power,[24] while some researchers report that a small interval with larger length is more desirable for noise reduction regardless of the form of the serrations.[24-26] It is also found that the sawtooth and wavy forms are more effective than others in noise reduction by the trailing-edge serrations.[22, 27] Recently, 3D features including the inclination (or skewness) of the serrations have been studied, resulting in better noise reduction capabilities by experiments[28] and a novel mechanism is discovered by CFD methods but lacks the support of acoustic analysis.[29] LE noise in airfoils can be reduced by the destructive interference of the scattered pressure or incoherent response along the span induced by the sawtooth or wavy serrations.[23, 30, 31] In addition, Rao et al. discovered that the suppression of KH instability by long and narrow slotted LE serrations contribute to the self-noise (TE noise) reduction of the airfoil, by modifying flows downstream of the serrations.[32] However, it remains insufficiently studied what the underlying mechanism of the morphological effects of the LE serrations on noise reduction, particularly in terms of the 3D features of the inclination, which may play a crucial role in improving the acoustic performance in rotary fluid machinery.

Despite these endeavors to explore the fundamental aeroacoustic mechanisms of serrated airfoils, the idealized 2D models cannot be employed directly for the design of industrial fans.[6] The surrounding flow field and noise signatures of the LE-serrated blades remain under-researched, particularly with regard to the morphological features.[12, 33-35] While a recent study on the morphological effects of serrated fans reports results similar to those of airfoils, longer lengths and smaller intervals are preferable,[35] the 3D features of the serrations, such as inclination or curvature, remains barely studied for rotary devices. Recently, Wang et al. reported the acoustic benefit of the serration inclination in a mixed flow fan. They found through large-eddy simulations (LES) that, the LE noise caused by flow separation can be suppressed using slotted

LE serrations that induce high local incidence angles; this, in turn, leads to the mitigation of stagnation and the reduction of KH instability.[12] This indicates that the LES-based high-resolution studies of the complex near-field flow structures involving laminar-turbulent transitions are critical to clarifying the mechanisms underlying the morphological effects of the serrations for the rotary blades.

In addition, aeroacoustic features have been studied in combination with computational fluid dynamics (CFD). Semi-analytical methods informed by CFD have been developed based on theories related to the self-noise of airfoils in recent decades.[36-38] To date, fast predictions for the broadband noise of a fan can be achieved based on Amiet's model with the statistical data for turbulent flows derived from the Reynolds-Averaged Navier–Stokes (RANS) simulations.[13, 39] Direct computational aeroacoustics (CAA) have also been used with the Lattice-Boltzmann Method to directly solve tight coupling problems associated with the unsteady and compressible flow and noise generation.[40, 41] However, given the high computational cost of the direct CAA, the hybrid method to combine the Ffowcs Williams-Hawkings (FWH) analogy with LES can be a reasonable choice. With this strategy, the FWH analogy can provide analytical solutions for the far-field sound radiating from a randomly moving body surface with the accurate surface pressure data obtained from the LES.[3, 42-44] With respect to the acoustic source identification, however the common method of monitoring wall pressure fluctuations cannot provide accurate information of the surface noise strength (SNS) for a specific location where the sound is received, as it does not consider the influence of the movement (relative to the receiver) and the geometry of the body surface incorporated in the FWH equation.[12, 40, 41, 45]

To explore the morphological effects of LE serrations on the acoustic signatures of rotary devices, we designed a series of slotted serrations in terms of three parameters, namely, the interval, length, and inclination angle, and appended them on the LEs of a mixed flow fan. Acoustic and aerodynamic measurements were carried out at an operating point and one-third octave spectral analyses were conducted for the collected sound data to analyze the variations in the acoustic signature. To our best knowledge, the LES-based FWH analogy is, for the first time, utilized to visualize the SNS on the rotor surface (section II.C). A simple index for the evaluation of separation severity was proposed based on the pressure gradient (PG) information derived from RANS analyses (section IV.A). We confirmed the robust tradeoff resolving capability of the serrations under various morphologies and clarified the acoustic effects of the three studied parameters. In addition, a new passive flow control mechanism was discovered that involves the inclination as well as its side effects. Our work can deepen the understanding of the morphological effects of the serrations on fan noise and provides researchers with useful methods in terms of noise identification and the fast evaluation of chordwise separation.

## II. MATERIALS AND METHODS

### A. Designs and experimental methods

We considered the same mixed flow fan used in our previous studies[12, 46] as the prototype, which as depicted in Fig 1(a) comprises an impeller having six blades and nine outlet guide vanes (OGVs) downstream. Parameters and a typical operating condition of the fan are summarized in Table 1. In this study, the LE-serrated blades and the basic design (i.e., the prototype) are studied. Fig. 1(b) shows the serrated impeller, whose LE serrations are bent slightly to adapt to the spanwise curvature of the LEs.

**TABLE I.** Parameters of the basic mixed flow fan

| Parameters | Values |
|---|---|
| Outlet radius of impeller $R_2$ (mm) | 213 |
| Outlet and inlet radius ratio $R_2 R_1^{-1}$ | 1.4 |
| Blade chord length at midspan (mm) | 194 |
| Stagger angle at blade root $\gamma$ (°) | 53 |
| Specific speed $n_s$ | 264 |
| Flow rate $Q$ (m³ min⁻¹) | 34.8 |
| Rotational speed $N$ (r min⁻¹) | 1431 |
| Reynolds number | 46000 |

Slotted-type LE serrations[22] based on a 1-mm thick ABS (Acrylonitrile-Butadiene-Styrene) flat plate are employed due to the ease of their fabrication. Morphological effects of LE serrations on the aeroacoustic characteristics of the mixed flow fan are investigated in this study in terms of three morphological parameters, namely the length, the inclination angle (toward the tip of the blade), and the interval (taken as double the serration width), as shown in Fig. 1(c). The fan designs with LE serrations are defined in terms of three parameters, such as "serL15A20I2" that indicates the combination of a length of 15 mm, inclination angle of 20 degrees, and an interval of 2 mm. For simplicity, we employ the term of "serL15A20" herein, as most fan designs share the same interval of 2 mm while designs featuring different intervals will be labeled with the three parameters. Besides, considering that LE serrations may extrude to the shroud (Fig. 1(a)), these must be trimmed off in some designs with a high inclination angle to ensure proper clearance at the blade tip.

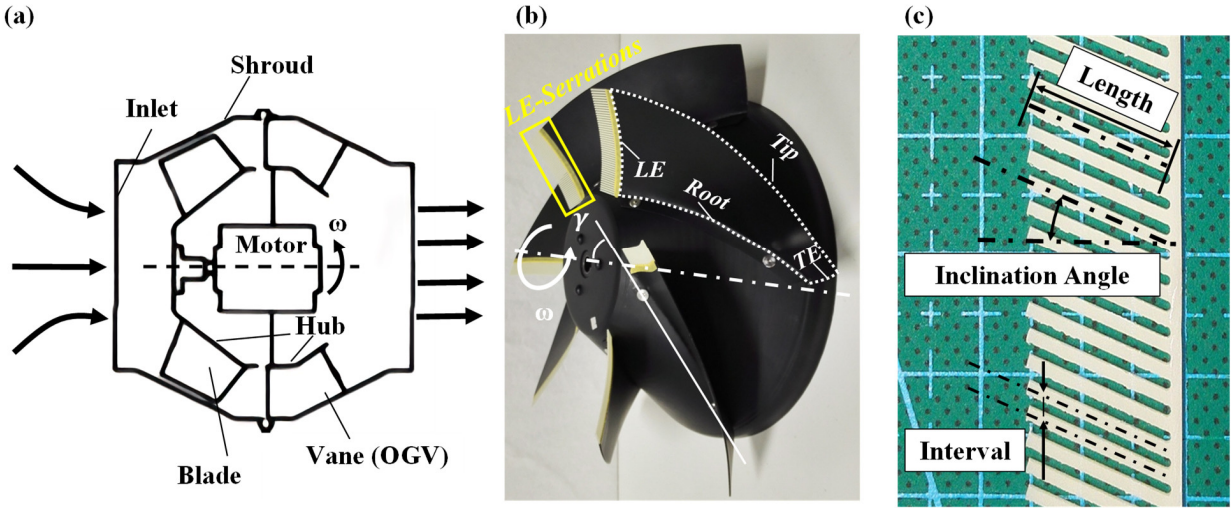

**FIG. 1.** (a) Configuration of the studied mixed flow fan, (b) LE-serrated impeller, and (c) three morphological parameters for slotted-type LE serrations.

Aeroacoustic and aerodynamic experiments undertaken in our previous studies[12, 46] were repeated using experimental set-ups depicted in Figs. 2 (a), (b). The aerodynamic performance was evaluated by computing the total pressure efficiency, such as:

$$\eta = \frac{(P_{out} - P_{in}) \cdot Q/60}{P_{input}}, \quad (1)$$

where $P_{out}$ and $P_{in}$ denote the total pressure at the outlet and inlet, respectively. $P_{out}$ is measured by a manometer (Fig. 2(a)) that considers the pressure loss based on JIS B8330; $P_{in}$ is taken as 0 Pa at the inlet which is open to the atmosphere; and $P_{input}$ is the input power of the motor measured by the power meter in Fig. 2(a).

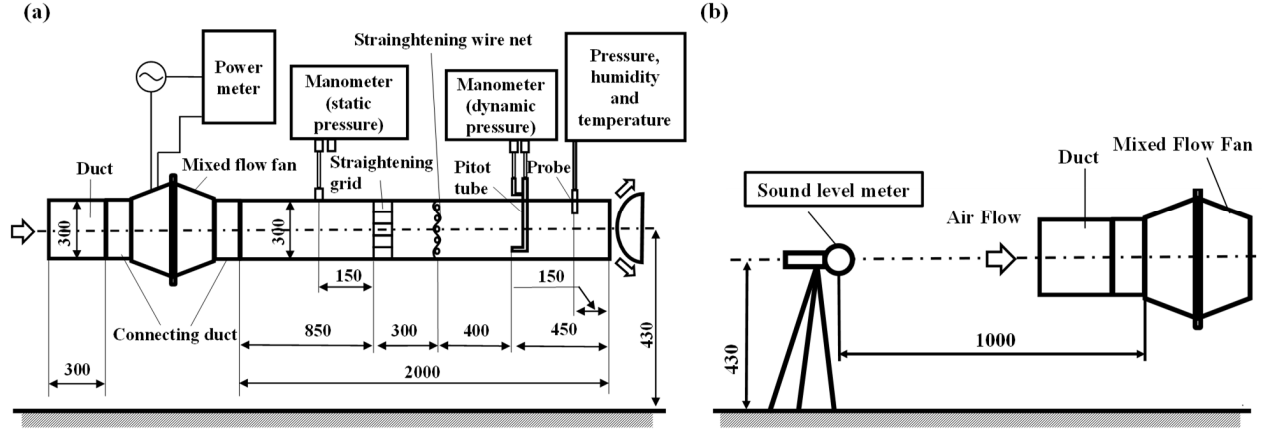

**FIG. 2.** Experimental set-ups for (a) aerodynamic performance measurement (by CC BY), and (b) acoustic performance measurement. Dimensions in mm.

Investigation of the aeroacoustic characteristics was conducted by collecting various noises using a sound level meter, which, as shown in Fig. 2(b), was mounted along the impeller axis 1 m away from the center of the section connecting the inlet duct in accordance with JIS B 8346. The noises were then analyzed using the one-third octave band.

To quantify and examine the aeroacoustic performance in association with different fan designs, we further employed the specific noise level (SNL) in addition to the sound pressure level (SPL), aiming to provide a fair comparison by rationalizing the experimental results obtained with slightly different rotation speeds for different designs,[12, 47]. The SNL is defined as,

$$L_s = L_{in} - 10 \log_{10}\left(\frac{\widetilde{P_{out}}^{2.5} Q}{60}\right), \qquad (2)$$

where $L_{in}$ denotes the SPL (dB) weighted by A scale, $\widetilde{P_{out}}$ is the standardized total pressure at the outlet, which is obtained conforming to JIS B8330 and given by,

$$\widetilde{P_{out}} = \frac{\rho_0}{\rho} \times P_{out}, \qquad (3)$$

where $\rho_0$ is the standardized density (1.2 kg m$^{-3}$) and $\rho$ is the density measured experimentally.

## B. Computational Fluid Dynamic Models

### 1. Two rotor-only models and grid systems

In this study, we focused on blade LE noise without considering the OGVs, and hence, employed a rotor-only (or rotor-alone) configuration model[1, 48] shown in Figs. 3 (a), which we herein term as Complete Blade Model (CBM). To further save the computational cost, a Partial Blade Model (PBM) was also employed for the large scale LES analyses of various morphological designs (Fig 3 (b)). PBM can be regarded as the rotational domain of the CBM sectioned by a plane vertical to the impeller's axis, with 20% chordwise reservation at the blade root and 46% at the tip due to the 3D curvature of the blade. According to our earlier study,[12] the flow separation causes no discernible APG beyond 10% blade-chord from the LE where the modifications by serrations are observed (also see in Fig. S1, supplementary material), and thus, the PBM is large enough to cover the crucial locations for pressure information . The commercial solver ANSYS 14.5 CFX is employed to conduct CFD analyses for this study.

The two models share the same boundary conditions (BCs) including a periodic BC that is one-sixth of the full cycle and have three parts, namely, the inlet duct region, rotational domain, and outlet duct region (Fig. 3(a)). The 'general grid interface' (GGI) model in CFX is employed to connect the three parts and the 'frozen-rotor model' is utilized to deal with the frame change at the two interfaces, as the rotational domain takes a rotational frame to avoid the mesh moving, while the other parts are in the steady or ground frame. The outlet duct has 1 m length that extends from the end of the rotational domain and thus, is not shown (in Fig. 3 (a) and (b)) for better view. The total inlet pressure is set as 0 Pa with vertical flow direction, while the mass flow rate at the outlet is selected according to the operating condition in Table I.

A local right-handed coordinate system was chosen to depict the flow field variations as in our earlier study.[12] The origin is at the root of LE as shown in Fig. 3 (c). The $X_1OX_2$ plane is determined by three points, $T_1$, $O$ and $T_2$; the $X_1$ direction is determined by line $OT_1$; and the interval between $T_2$ and $T_1$ is taken as 10% chordwise length. The shaded area is defined as the LE region and thus, the $X_3$ direction is considered normal to it while $X_2$ can be regarded vertical to the LE.

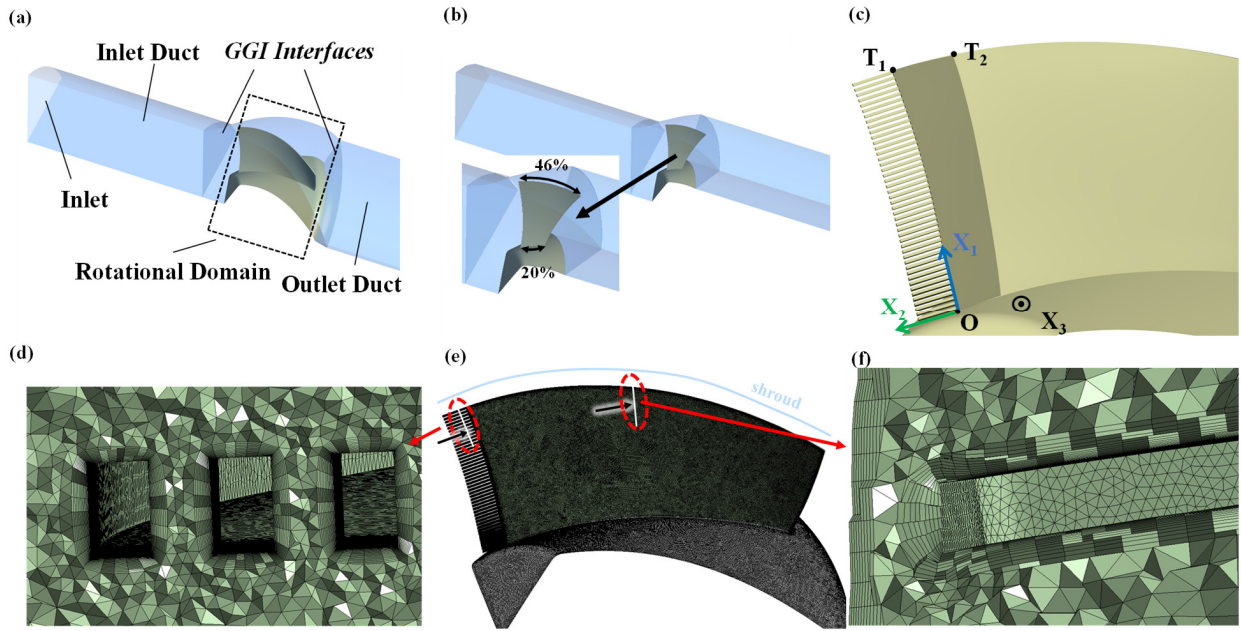

**FIG. 3.** CFD model for the (a) CBM and (b) PBM; (c) the definition of the local coordinate system and the LE region; grids (of serL15A0) at (d) the tip serrations, (e) blade surface and (f) shroud.

The mesh density and prism layers of the grid systems, particularly for the crucial rotational domain, have been chosen from existing literature which has been already verified and validated.[12] As seen in Figs. 3 (d) – (f), the unstructured grids constructed using ANSYS Meshing Module 14.5 enable easier adaptation to the complex morphology and geometry of the blades, and at least five nodes were imposed on each serration width-wise as in available literature.[32] Mesh Set 2 in Table II is used for this study and it is suitable for the LES analyses with 15 prism layers and y+ = 1. PBM follows the same grid settings as the CBM and thus the description is omitted for conciseness. More details are seen in section II.B.3.

**2. Turbulence modeling and solution settings**

Both steady RANS and unsteady LES analyses are conducted based on the assumption that the fluid is incompressible. The former can quickly provide primary information regarding near-field flow features at a low cost by overlooking irrelevant transient details, which makes the comparative evaluation of serration morphology cost-effective. Despite their high computational cost, LES analyses can reproduce accurate flow fields in terms of the intense laminar-turbulent transitions around the blades, which is necessary for the implementation of the FWH analogy.

Therefore, steady RANS analyses are carried out using the CBM for all designs to investigate PG variations in the LE region while LES analyses mainly use the PBM for the basic design, serL15A0, and serL15A40 to investigate the effects of serration inclination. For the basic design, the CBM is also used for LES analysis. However, the design with the largest inclination angle, serL15A60, is not analyzed using LES as the computational cost was very high, presumably due to the complex blade-surrounding flow field.

## 2.1 Steady RANS analyses

The governing equation of continuity and momentum of the incompressible and steady RANS can be written as:

$$\frac{\partial \bar{u}_i}{\partial x_i} = 0, \tag{4}$$

$$\overline{u_j}\frac{\partial \bar{u}_i}{\partial x_j} = -\frac{1}{\rho}\frac{\partial \bar{p}}{\partial x_i} + \frac{\mu}{\rho}\frac{\partial^2 \bar{u}_i}{\partial x_j \partial x_j} - \frac{\partial}{\partial x_j}\left(\overline{u'_i u'_j}\right). \tag{5}$$

The shear stress transport (SST) turbulence model[49] is employed here along with the high resolution advection scheme. The 'physical time scale' of the false time step or iteration step is set to $1/\omega_r$ where $\omega_r$ denotes the rotational speed, as recommended by CFX. To obtain high-accuracy results related to the boundary layer, a root mean square (RMS) residual $< 1 \times 10^{-6}$ is adopted as the convergence criterion while cases that cannot satisfy this criterion are considered acceptable if their torque fluctuation is within 0.01% for the last 200 iterations.

## 2.2 LES analyses

For the LES analyses, the continuity equation is given in Eq. 4, while the momentum is governed by the incompressible Navier-Stokes equation

$$\frac{\partial \bar{u}_i}{\partial t} + \frac{\partial (\bar{u}_i \bar{u}_j)}{\partial x_i} = -\frac{1}{\rho}\frac{\partial \bar{p}}{\partial x_i} + \frac{\partial}{\partial x_j}\left[\frac{\mu}{\rho}\left(\frac{\partial \bar{u}_i}{\partial x_j} + \frac{\partial \bar{u}_j}{\partial x_i}\right)\right] - \frac{\partial \tau_{ij}}{\partial x_j}, \tag{6}$$

where $\tau_{ij}$ denotes the subgrid-scale (SGS) stress tensor, and the adopted eddy-viscosity assumption is given by,

$$\tau_{ij} = \frac{\delta_{ij}}{3}\tau_{kk} - 2v_t \bar{S}_{ij} = \frac{\delta_{ij}}{3}\tau_{kk} - v_t\left(\frac{\partial \bar{u}_i}{\partial x_j} + \frac{\partial \bar{u}_j}{\partial x_i}\right), \qquad (7)$$

$\delta_{ij}$ is the Kronecker symbol while $v_t$ denotes the turbulent eddy viscosity. The wall-adapted local eddy-viscosity (WALE) model proposed by Nicoud and Ducros is used to compute the SGS viscosity as in the most turbomachinery noise studies.[1, 3, 50] Central difference and second-order backward Euler were adopted for the advection and temporal schemes, respectively, with the RMS residual $< 5 \times 10^{-6}$ for each time step.

LES computations are initialized using the corresponding results from steady analyses. The time steps of the four LES analyses are varied to adapt to the different average mesh density in each case, such that CFL falls between 0.5 and 1 as suggested by CFX. Thus, 2.5e-5 s, 1e-5 s, and 5e-6 s are chosen for the CBM of basic design, PBM of basic design, and other two PBMs (serL15A0 & serrL1540), respectively. Results during the period from 0.1328 s – 0.1747 s, corresponding to $3.17T - 4.17T$ or $19\,T_b - 25\,T_b$ (where $T$ is the elapsed time for a complete revolution and $T_b = T/6$, denoting the time of passing for one blade), are used for the FWH analogy. Results for the basic design case using CBM are saved for each step (i.e., 2.5 e-5s), while for the other three cases using PBM, results are saved every 2e-5s, corresponding to every 2 or 4 steps.

## 3. Verifications

Verifications are still carried out in this research to (1) check the grid independence for a RANS-informed index *SAPG* which will be introduced in section IV.A; (2) rule out the influence of initial conditions for LES analyses, although the grid systems and solution settings have already been verified in literature.[12]

The grid independence study in Table II was conducted based on steady CFD analyses, where the fine Mesh 3, medium-fine Mesh 2, and relatively coarse Mesh Set 1 are listed in different columns. Mesh Set 2 is found able to reproduce the same impeller torque and close *SAPG* (1.11% error) compared with the Mesh Set 3. Fig. S1 also shows the PG on the suction faces of the three mesh sets, where the Mesh Set 2 shows a close spatial distribution (which matters for *SAPG*) to the Mesh Set 3, indicating a satisfying accuracy for this study.

**TABLE II.** Grid independence study for CBM

| Parameters & Results | Mesh Set 1 | Mesh Set 2 | Mesh Set 3 |
|---|---|---|---|
| Total node number ($\times 10^6$) | 2.23 | 3.05 | 3.88 |
| Blade surface grid size (mm) | 1 | 0.75 | 0.75 |
| Rotational domain grid size (mm) | 3.5 | 2.5 | 1.5 |
| y+ on blade (based on Re= 450000) | 2 | 1 | 1 |
| Growth rate of prism layer (above blade) | 1.2 | 1.2 | 1.2 |
| Prism layer number (above blade) | 12 | 15 | 18 |
| Impeller Torque (N·m) | 1.752 (0.29%) | 1.747 (0%) | 1.747 |
| *SAPG* of LE region (kg· s$^{-2}$) | 41.98(1.15%) | 41.96(1.11%) | 41.50 |

For the LES analyses, the moving-averaged torque (MAT) for one blade passing ($T_b$) is calculated to examine whether the time window ($0 - 3.17T$) is large enough to overcome setup sensitivity (Fig. S2, supplementary material). Theoretically, the simulated MAT value should converge as the number of revolution increases. The standard deviations and maximum fluctuations of MAT are all below 0.9% and 1.8% of their mean values for each LES case considered, for $3.17T - 4.17T$, suggesting that the simulated flow fields are independent of the initial setup conditions.

## C. Noise source analyses via aeroacoustic computation

### 1. Ffowcs Williams–Hawkings analogy and verification

The Ffowcs Williams–Hawkings (FWH) equation is employed for aeroacoustic analyses combined with LES simulations. Several forms of the FWH equation were derived by F. Farassat under the assumptions of negligible quadruple noise and low Mach-number viscous stress.[51] The form below (i.e., the equation (9) of Ref. 51 ) is employed herein for convenience as the integrations can be carried out directly on the body surface,

$$p_f(\vec{x},t) = \frac{1}{4\pi}[\frac{1}{c}\frac{\partial}{\partial t}\int_{S^*}(\frac{\rho c v_n + p_s cos\theta}{r|1-M_r|})_{\tau^*}dS + \int_{S^*}(\frac{p_s cos\theta}{r^2|1-M_r|})_{\tau^*}dS] \qquad (8)$$

where $S^*$ denotes the body surface when no parts of the body move faster than the sound; $p_f(\vec{x},t)$ is the far-field acoustic pressure at the sound receiver location ($\vec{x}$) at the instant of $t$, while $\vec{r}$ is the vector pointing to the receiver from a certain point on the surface ($r = |\vec{r}|$). $\tau^*$, $p_s$, and $c$ are the sound emission time, surface pressure, and sound speed, respectively. It is worth noting that $p_s$ and $p_f$ are both expressed as $p$ in Farassat's work, but here we differentiate them because $p_s$ is derived from CFD results while the $p_f(\vec{x},t)$ is as the solution to Eq.8. Given $\vec{n}$ as the normal vector from the body surface, the angle between $\vec{n}$ and $\vec{r}$ can be defined as $\theta$, while $v_n$ and $v_r$ represent the velocity components in the directions of $\vec{n}$ and $\vec{r}$, respectively. Here, we solve the acoustic pressure corresponding to the sound collection location chosen in the experiments (Fig. 2(b)), which is at the impeller axis corresponding to a zero $M_r$ defined by $v_r/c$ as $v_r$ is 0. The two sides of this equation use different time references due to the retardation of sound propagation equal to $r/c$.[51]

The variables on the right-hand side (RHS) of Eq. 8 are derived from the LES analyses. The CFX solver adopts the element-based finite volume method, which computes and stores all the flow-related variables including pressure and velocities at the nodes or mesh vertices.[52] Thus, the RHS variables, such as the blade surface pressure, in Eq.8 can be obtained by directly exporting node-based data from ANSYS CFD-Post.

We further developed an in-house MATLAB code for solving the FWH analogy, which is verified via analytical method for the special receiver location selected in this study (Fig. 2(b)). It is worth noting that validating the code is difficult because, (1) our rotor-only model neglects the OGVs which are unavoidable components of the experimental set-up; (2) the FWH analogy overlooks the refraction and reflection of the sound wave due to the shroud which is also an unavoidable component for this study. Fortunately, the location of the receiver on the revolving axis and the constant revolving rate of the impeller enable the verification because this unique situation simplifies the RHS of Eq.8 considerably. Details of the verification are summarized in Appendix, and the current code can limit the maximum error to less than 0.7% compared to the analytical result when the time step number exceeds 1500.

**2. Identification of noise sources**

Herein, we propose a novel method to visualize the surface noise strength (SNS) based on the FWH analogy computations to identify noise sources and discover their distribution characteristics. The SNS can be defined as $dp_f^2/dS$ where the logarithmized metrics of $p_f^2$ is adopted to measure the sound pressure level (SPL), which Siddon solved by cross-correlating the $p_f$ and $p_s$ values measured experimentally to identify surface dipole sources;[53] but we herein derived a formula by solving Eq.8, such as:

$$\frac{dp_f^2}{dS}(\vec{x},t,\vec{y}) = 2p_f(\vec{x},t) \cdot \frac{dp_f}{dS}(\vec{x},t,\vec{y}) = \frac{p_f}{2\pi}\left[\frac{1}{c}\frac{\partial}{\partial t}\left(\frac{\rho_0 c v_n + p_s \cos\theta}{r|1-M_r|}\right)_{\tau^*} + \left(\frac{p_s \cos\theta}{r^2|1-M_r|}\right)_{\tau^*}\right] \quad (9)$$

$p_f(\vec{x},t)$ denotes the transient sound pressure at a specific receiver ($\vec{x}$), an exact solution to Eq.8, while $dp_f/dS(\vec{x},t,\vec{y})$ represents the derivative of $p_f(\vec{x},t)$ relative to the body surface area $S$ at a certain position ($\vec{y}$) on the body surface (i.e., $\vec{y} = \vec{x} - \vec{r}$), which thus stands for the local contribution to $p_f$ of the body surface. The $dp_f/dS$ is obtained by solving Eq.8 numerically before carrying out the integration over the wall surface. Therefore, for a specific receiver with fixed $\vec{x}$, one can obtain the time-varying $dp_f^2/dS$ at any point on the body surface via interpolation, thus allowing us to conduct the SNS visualization over the entire wall surface. Note that the unique receiver location at the revolving axis leads to $\partial v_n/\partial t = 0$ (see Appendix for details), and thus, the SNS in this study contains only the dipole component.

It is known that the application of FWH may overpredict the noise level for a ducted fan as this approach does not consider the refraction and reflection due to the existence of, e.g., a casing or shroud.[42] Here, we aim to clarify the noise production feature irrespective of the sound propagation-related issues such as cut-off modes within the duct for which the FWH analogy can be an effective method in resolving the surface noise with Eq.9 at a certain specific receiver. Acoustic signatures at some locations are crucial to engineering applications, like the location assigned for testing by JIS B 8346 (Fig.2(b)). By contrast, the wall pressure fluctuations that are often monitored to analyze noise sources generally consider neither the location of the sound receiver nor the shape of the body surface.[12, 40, 41, 45] Therefore, the current SNS visualization method is capable of quantifying and visualizing the local acoustic characteristics more specifically and accurately, rendering it helpful for researchers to identify the noise sources.

## III. RESULTS

### A. Overview

LE-serrated airfoils generally confront a tradeoff problem that the aerodynamic performance may be harmed while the owl-inspired LE serrations suppress the KH instability.[32, 54] Nevertheless, earlier studies have confirmed that LE-serrated airfoils can generate competitive aerodynamic performance compared to unserrated ones at high angles of attack (AoA) $\alpha_g$ (over 15°).[25, 55] Thus, the morphological effects of the serrations must be studied to exploit this tradeoff-resolving capability.

The results of our experiment reveal that the tradeoff issue is robustly resolved regardless of the morphological variations of the serrations for the mixed flow fan. Fig. 4 presents the results of 10 × ΔSpecific Noise Level vs. ΔEfficiency at the operating point (34.8 m$^3$ min$^{-1}$). ΔSpecific Noise Level is calculated by subtracting the SNL of the biomimetic design from that of the basic design while ΔEfficiency can be calculated by subtracting the efficiency of the basic design from the serrated design. Hence, the result of the basic design is located at the origin, and the first quadrant represents both the increase in efficiency and the noise reduction. The best design is serL15A20, which achieves a noise reduction of 1.4 dB corresponding to a 1.5% increase in efficiency. Nearly all LE-serrated designs cluster in the first quadrant except for serL5A0 and serL15A60, demonstrating the tradeoff resolving ability of the serrations under different morphological parameters. The efficiency of serL5A0 is merely 0.2 percentage point lower than the basic design, while serL15A60 generates a marginally higher noise of 0.02 dB. The successful tradeoff resolution should be ascribed to the large stagger angle (Table I) of this mixed flow fan, corresponding to a large blade angle (akin to geometric AoA) at LE, which is approximately 40º at the blade midspan. This suggests that the aerodynamic performance can be barely affected by the LE serrations regardless of the morphological variation, although an improper design parameter may also counteract the noise reduction benefits of the serrations.

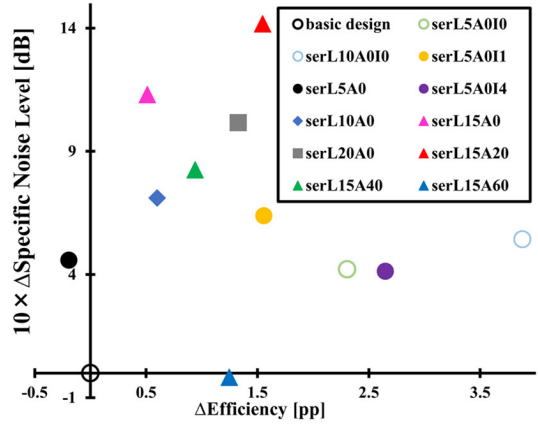

**FIG. 4.** Specific noise level (10 × ΔSpecific Noise Level) vs. efficiency (ΔEfficiency) in a manner of differences from the basic design at the operating point.

As shown in Fig. 4, the designs of serL5A0I0 and serL10A0I0 are also included in considering the morphological effects compared to the remaining optimization space. Note that serL5A0I0 and serL10A0I0 are the designs appended with the extended plate (with a zero interval) at LEs despite the label of 'ser'. Experiments using designs with extensions exceeding 10 mm were not carried out for safety reasons. Both serL5A0I0 and serL10A0I0 outperform the basic design in aerodynamics and aeroacoustics, indicating that the prototype has room for optimization; whereas they perform worse than serL5A0I1 and serL10A0 in sound reduction by 0.22 and 0.17 dB, respectively, suggesting that the aeroacoustic modifications by serrations cannot be attributed to the remaining optimization space. Moreover, technically, the flow field variations induced by serrations are entirely different from those with the extended plates.[12, 32] Therefore, the performance improvement with LE serrations is mainly due to the flow modifications induced by their own morphologies, and the effects of the remaining optimization space should be excluded, particularly for the aeroacoustic variations.

Fig. S3 (a)-(c) in supplementary material indicate the sound pressure level (SPL) of the three LE-serrated blade groups vs. varied design parameters derived from our acoustic measurements without any correction (i.e., $L_{in}$ in Eq.2). Compared to the SNL values in Fig. 4, the noise level of serL5A0I1 in Fig. S3 (a) is slightly underestimated, but the sound level trends under the variation of each parameter are still consistent.

## B. Pressure gradient variations

Steady CFD analyses are carried out for all designs with CBM to explore the PG variations on the suction face caused by the LE serrations. Fig. 5 demonstrates typical PG modifications by serrations in the $X_1$ and $X_2$ directions (Fig. 3 (c)) with the basic design and serL15A20 as examples, which are hereafter termed as $PG_1$ and $PG_2$, respectively. The near-fields of both the designs were thoroughly studied using LES analyses (involving OGVs) in our previous research as well.[12]

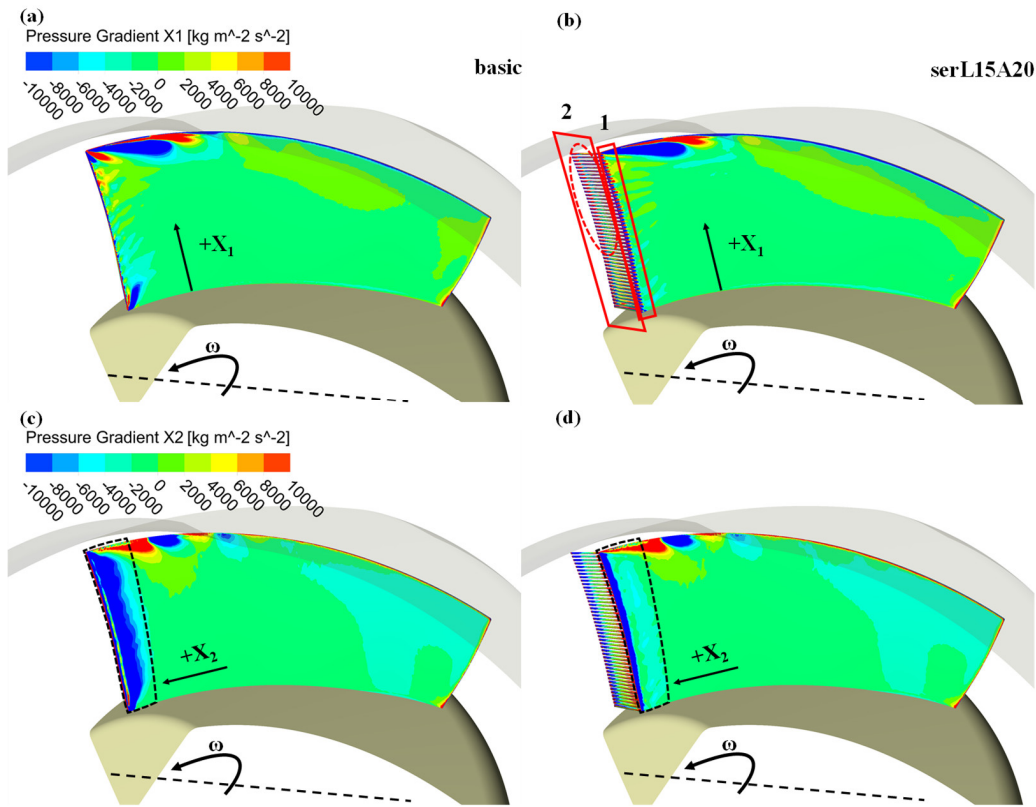

**FIG. 5.** Pressure gradient in $X_1$ and $X_2$ directions ($PG_1$ and $PG_2$), for (a) & (c) basic design, (b) & (d) serL15A40, respectively.

The PG$_1$ are found to be related to the by-product of high-frequency noise induced by the LE serrations. As seen from the Figs. 5 (a) vs. (b), LE serrations cause small PG$_1$ (negative and positive) pairs to concentrate at the base of the serrations (marked by 1), while the suction face also bears intensive PG$_1$ (marked by 2). Moreover, the tip-towards increase trend of PG$_1$ on the serrations (dashed circle in 2) is consistent with our previous findings regarding high-frequency fluctuations. We had concluded that serrations cause high-frequency fluctuations at these two positions from the small eddies of broken-up vortical flows,[12] and thus, it is the intense PG$_1$ found here that drives the generation of these eddies.

The remarkable modifications of the negative PG$_2$ (APG) by the serrations suggest that the flow separation was suppressed considerably. We had previously discovered that the transient PG$_2$ pairs of favorable PG and APG above the LE region were remarkably alleviated by LE serrations, resulting in the reduction of the separation noise at 40–4k Hz for this fan,[12] while the steady RANS results indicate that the dominant APG region in the LE decreased noticeably with serrations (Figs 5.(c) and (d)). Note that intensive APG can still be observed near the LE for serL15A20, indicating an interesting and typical pattern of PG$_2$ modification in which the APG region shrinks towards the LE. In section IV.A, an evaluation method for separation severity or separation-induced noise will be proposed based on this observation by quantifying APG spatial features.

## C. Surface noise strength

SNS is visualized by solving Eq.9 based on recorded data from LES analyses to figure out the noise contributions from the blade surface. The solution of $dp_f{}^2/dS$ for a complete revolution ($3.17T - 4.17T$) is selected to compute the time-averaged SNS on different blade suction surfaces (Figs. 6(a), (c)-(d)). Fig. 6 (b) is the near-field vortical structures in terms of the normalized helicity for the basic design. Normalized helicity is defined as $\vec{u} \cdot \vec{\omega}/(|u| \cdot |\omega|)$, where $\vec{u}$ and $\vec{\omega}$ are the flow velocity and vorticity so that movement of the vortices relative to their spin directions can be visualized.

Regions of negative SNS like the blade trailing edge in Figs. 6(a) do not necessarily imply noise reduction capability, but indicate that they are not the main noise sources for the computed receiver. This is because they could still generate noise during specific portions of a cycle or for other receiving locations, despite their total negative contribution. This may explain earlier failures of the biomimetic TE serrations or fringes for this fan.[12]

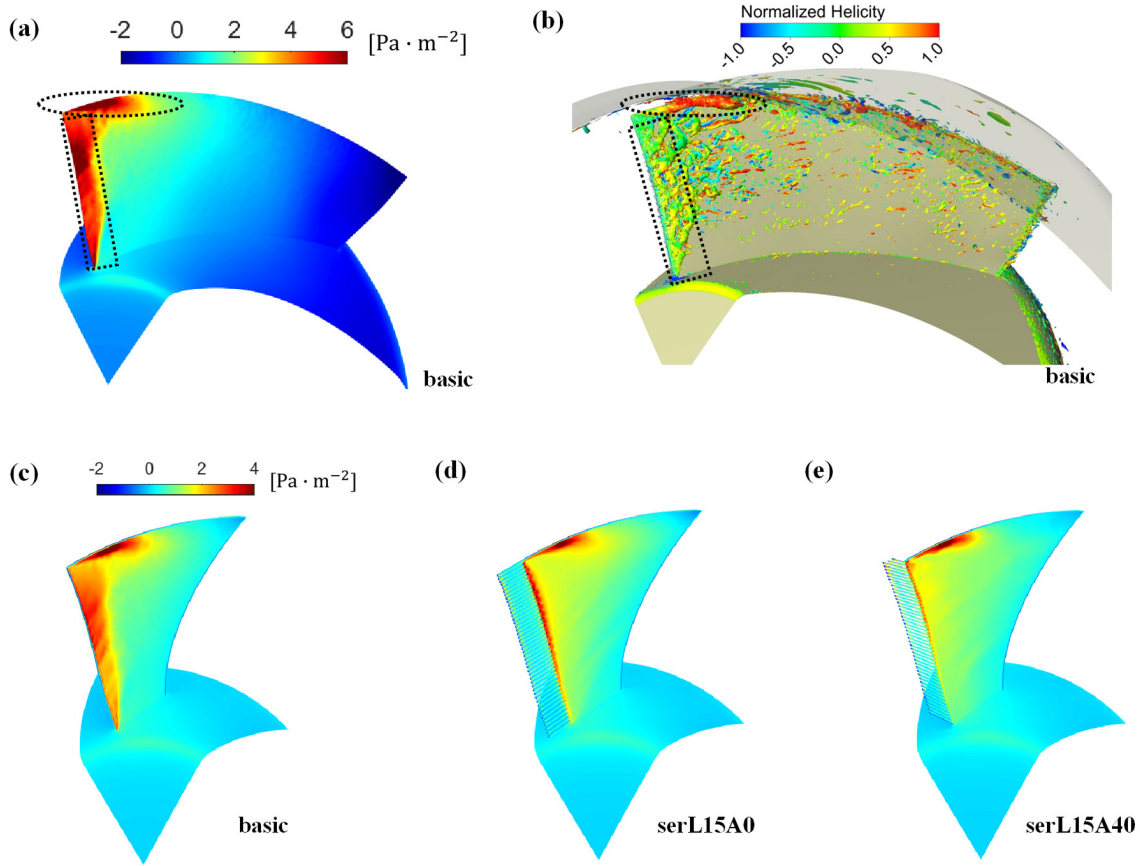

**FIG. 6.** Basic design results with CBM of (a) time-averaged surface noise strength (3.17T – 4.17T), and (b) iso-vorticity surfaces (instant of 3.17T or 0.15652s, swirling strength = 3000 s$^{-1}$) in terms of normalized helicity; time-averaged surface noise strength using PBM (3.17T – 4.17T) for (c) basic design, (d) serL15A0 and (e) serL15A40. All of these are based on LES analyses.

The regions of high positive SNS, however, suggest a strong net sound power output towards the receiver and can be reasonably suspected to be the main noise sources. As seen in the Figs. 6(a), the LE region and the tip manifest a powerful noise output (the dashed rectangle and circle), while severe flow spiraling and vortex shedding can be observed here as well (Figs. 6(b)). The SNS on the pressure face is omitted herein as there are no positions manifesting such intensive sound radiation as the above two. This suggests that the separation-induced LE noise and the tip noise induced by tip vortex forming through the clearance are probably the main noise sources, as these vortical structures can develop turbulence due to KH instability which then impinges onto the blade suction surface and generates sound.

Computational results with the PBM show consistent distribution of acoustic sources on the blade surface as well. As shown in Fig. 6 (c), the basic design shows nearly the same SNS distribution as that in Fig. 6 (a) though the overall level is lower due to the total decrease in noise with blade cutting, suggesting that it is reasonable to enable lower-cost explorations with PBM. Figs. 6 (c) vs. (d) or (e) reproduces the reduction in LE noise due to the suppression of KH instability by the serrations pointed out by the previous studies.[12, 32] The SNS distribution is quite similar between serL15A0 and serL15A40 (Figs. 6 (d) and (e)) in which the LE noise gets reduced while the tip noise is still obvious. However, serrations with large inclinations (≥ 40°) can deteriorate acoustic performance, which the time-averaged value fails to reflect and will be discussed in sections IV.B&C.

## IV. DISCUSSION

### A. Index for separation severity evaluation

Inspired by the correlations of the PG with noise generation (section III.B), we propose a fast evaluation method for separation severity or even separation-induced LE noise for the biomimetic design of this mixed flow fan. This index employs the $PG_2$ data derived from low-cost steady analyses with CBM, termed as the spatial sum of absolute pressure gradient (*SAPG*) and defined as,

$$SAPG = \beta \cdot IAPG = \beta \cdot \iint_{LE\ Region} |\nabla P_2| dS, \quad (10)$$

where $IAPG$ is the integral of the absolute pressure gradient and $\nabla P_2$ is exactly the $PG_2$. The absolute value ($|\nabla P_2|$) is taken to make the integrated variable positive and capable of reflecting the severity of APG in LE region without being counteracted by positive PG₂ despite slight overestimations. The nondimensional term $\beta$ stands for the 'spatial evenness' defined as,

$$\beta = 1/(\frac{\iint_{LE\ Region}\sqrt{(|\nabla P_2| - \overline{|\nabla P_2|})^2}dS}{IAPG}) = \frac{\iint_{LE\ Region}|\nabla P_2|dS,}{\iint_{LE\ Region}\sqrt{(|\nabla P_2| - \overline{|\nabla P_2|})^2}dS}, \qquad (11)$$

$$\overline{|\nabla P_2|} = IAPG/S_{LE} \qquad (12)$$

where the $\overline{|\nabla P_2|}$ is the average absolute PG₂ of the LE region and $S_{LE}$ is the area value of LE region which is a constant. $\iint_{LE\ Region}\sqrt{(|\nabla P_2| - \overline{|\nabla P_2|})^2}dS$ is akin to standard deviation of $|\nabla P_2|$ in the LE region and then normalized by $IAPG$ before calculating the inverse. Hence, a larger $\beta$ denotes a more even APG distribution, indicating that APG occupies more of the LE region given identical $IAPG$, which may also mean that APG affects the flows farther in the chordwise direction from LE in this research. The $SAPG$ is normalized as,

$$\widehat{SAPG} = \frac{SAPG}{SAPG_{|basic}} \times 100\%. \qquad (13)$$

$SAPG$ involves both the intensity and the spatial features of the APG. In Eq. 10, $\beta$ is multiplied as a factor as it can reflect the spatial effect of PG₂. Examples in Fig. 7 clarify the importance of $\beta$ by bringing out the mismatch in the results of experiments and $IAPG$. Fig. 7 (a) shows the sound pressure (mPa) differences (subtracting basic design' value) vs. frequency obtained from the one-third octave analyses, indicating that serL15A0 reduces the separation noise best (circled range over 100-3k Hz) while serL5A0I4 shows the least reduction. However, serL5A0I4's $IAPG$ in Fig. 7 (b) is quite close to that of serL15A0 and lower than serL10A0, leading to the wrong conclusion that the serL5A0I4 is prone to close separation severity compared to serL15A0, which ranks lower than serL10A0. On the other hand, Fig. 7 (c) and (d), the contours of PG₂ for the serL5A0I4 and serL10A0, successfully indicate that serL10A0 reduces the more area of severe APG in the LE region. Therefore, it is critical to involve the APG spatial characteristics to evaluate the separation severity and hence, $\beta$ in Eq. 10 is necessary. Consequently, with the involvement of $\beta$, $\widehat{SAPG}$ can reflect the reasonable conditions as in Fig. 7(a) for the separation severity (or noise) (Fig.8).

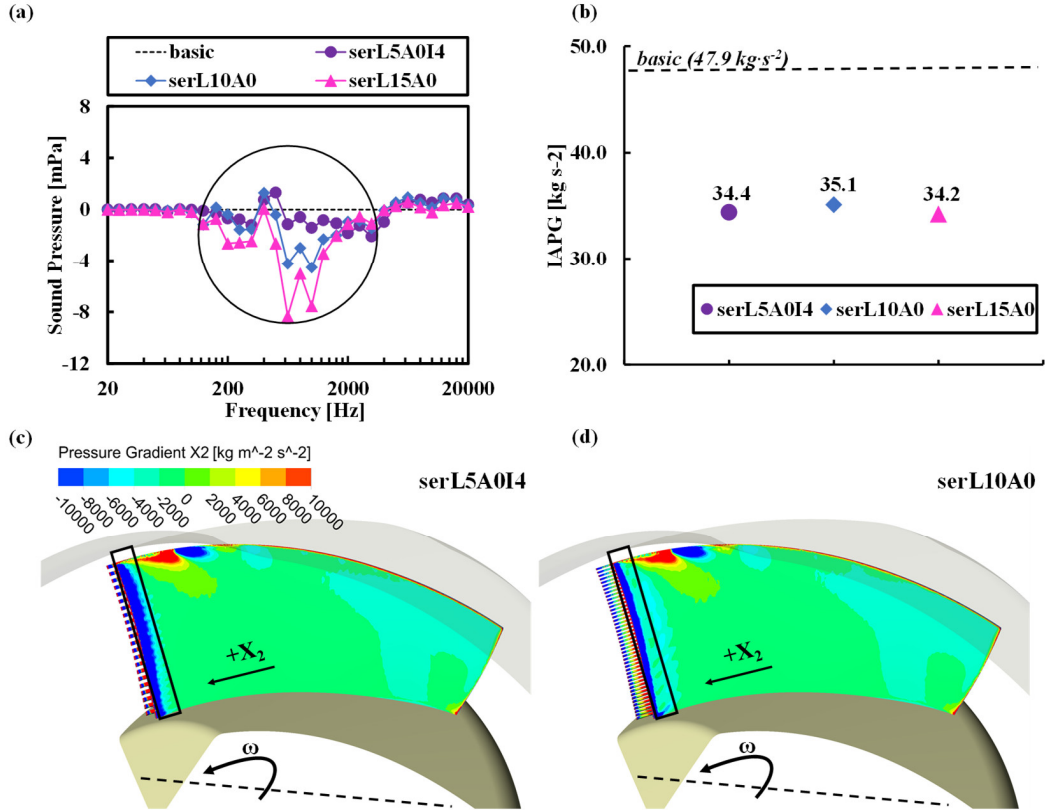

**FIG. 7.** Conflicts between results of (a) sound pressure differences (the basic design's value subtracted from three serrated blades) vs. frequency by one-third octave analyses of experimental data, and (b) corresponding *IAPG* results; PG$_2$ on the suction face for (c) serL5A0, and (d) serL10A0.

## B. Morphological effects on aeroacoustics

As seen in Figs. 8(a)-(c), one-third octave analyses are conducted for the sound data measured in experiments (with OGVs involved) and the basic design measurements are subtracted for comparisons. The results are arranged in three groups with certain varying parameters, i.e., interval, length, and inclination angle. Sound pressure in mPa is used instead of the sound pressure level to compare the designs in terms of acoustic features more clearly. Figs. 8(d)-8(f) show the corresponding results of $\widehat{SAPG}$ and $\beta$ ( *IAPG* is omitted for simplicity), which facilitates a comprehensive understanding of APG variations. Generally, as observed from Figs. 8(a)-(c), high frequency noise (> 6k Hz) generated by the serrations cannot be eliminated completely, while the separation noise is obviously reduced within 100 –3k Hz (see black circles), which is consistent with the above results related to LE-noise reduction (Figs. 6 (c) vs. (d) or (e)).

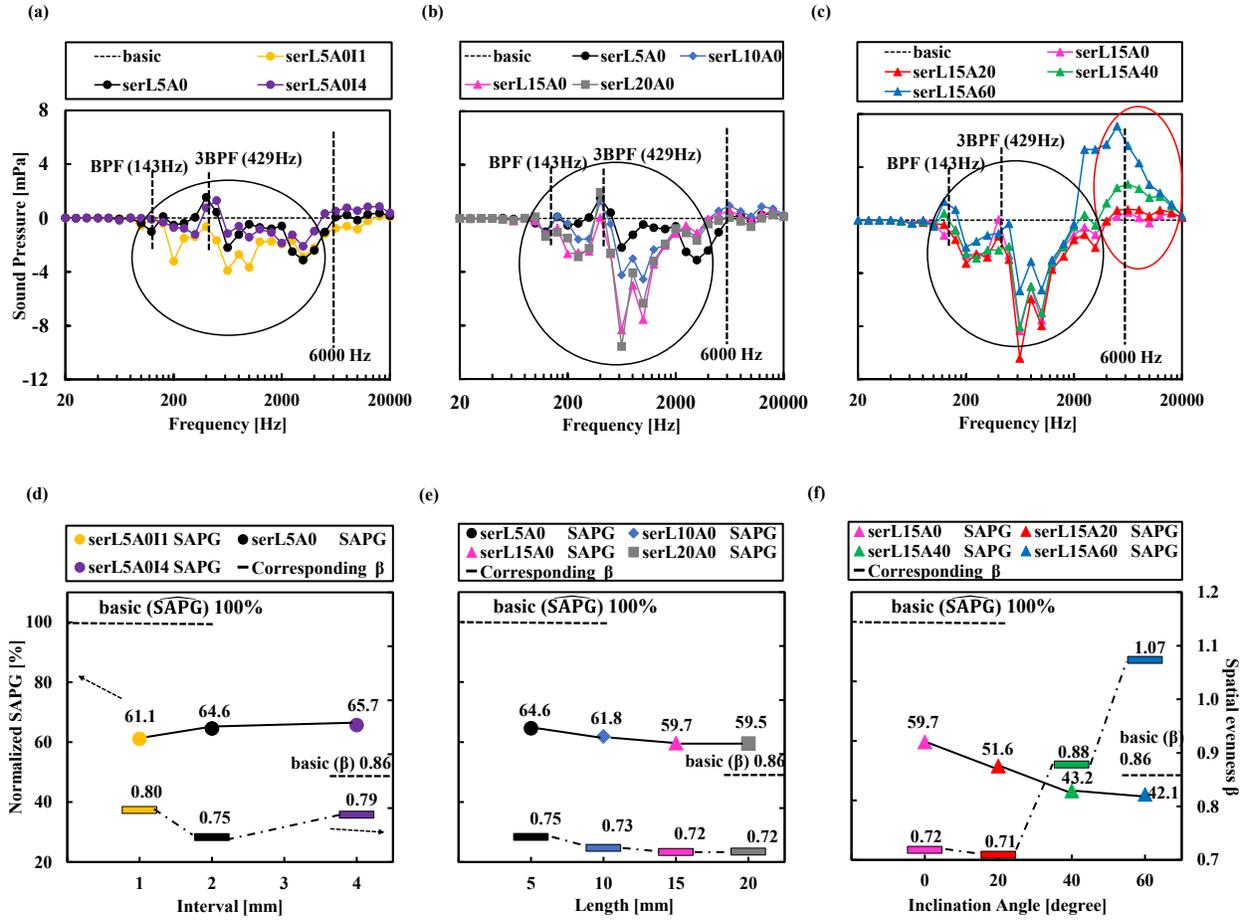

**FIG. 8.** One-third octave analyses of the sound pressure (experimental) for LE-serrated designs of varied (a) intervals, (b) lengths, and (c) inclination angles; $\widehat{SAPG}$ & $\beta$ (derived from steady analyses with CBM) for LE-serrated designs of varied (d) intervals, (e) lengths, (f) inclination angles.

Serrations with smaller intervals can result in the reduction of broadband noise as well as tonal noise. As seen from Fig 8(a), the shrinking of the interval can result in broadband noise reduction over a broad frequency range (100–20k Hz). The peak at 3BPF (blade passing frequency) can be lowered with a smaller interval as well, which is the smallest frequency for the tonal noise of blade-vane interaction and also the noise peak locates at for this fan.[12, 56] Fig. 8 (d) suggests that serL5A0I1 suppresses flow separation the most since it has the lowest $\widehat{SAPG}$ despite the highest spatial evenness $\beta$ (0.80). By contrast, serL5A0 has the highest $IAPG$ among them but with the lowest $\beta$ (0.75), and its $\widehat{SAPG}$ is still lower than serL5A0I4.

Hence, the interval affects the APG intensity and spatial distribution in a complex way; however, in general, there is less $\widehat{SAPG}$ with smaller size. Therefore, a smaller interval is more desirable, which can also be observed in Fig. 4 or Fig. S3 (a), consistent with the previous studies for airfoils.[25, 26]

A larger length can benefit the noise reduction significantly over a broad frequency range as well, but an optimum length exists for overall noise reduction. Fig. 8(b) demonstrates that the increase in length can effectively facilitate broadband noise reduction (beyond interval shrinking), particularly at frequencies of 100-3k Hz, while it marginally reduces the sound at frequencies in excess of 6000 Hz when the length is more than 10 mm. However, there seems to be a bottleneck for the reduction in separation noise (100-3k Hz), as further reduction is marginal when the length increases over 15 mm. Similarly, $\beta$ (from 0.75 to 0.72 in Fig. 8(e)) and $IAPG$ decreases along with the length extending, resulting in the decrease of $\widehat{SAPG}$ from 64.6% to 59.5%. However, further reduction is nearly impossible when the length exceeds 15 mm. This indicates that longer serrations can better modify the APG in the LE region, resulting in a pronounced decrease in separation noise, but with a limitation. On the other hand, an optimum length is observed for the tonal noise, which is approximately 15 mm as the sound pressure of serL20A0 at 3 BPF rises noticeably compared to the lowest serL15A0 (Fig. 8(b)). The fan-self tonal noise is mainly produced by force- and volume-displacement effects exerted by the rotating blades on the fluid,[4] and larger serrations can enhance these effects with the air, even though more flows may get controlled ahead of LE. Therefore, there exists a length optimum for overall noise reduction, which is approximately15 mm in our study and can also be easily observed in Fig. 4 or Fig. S3 (b).

Slight inclination ($\leq 20°$) of LE serrations can further help reduce noise with significant APG modification. Fig. 8(c) indicates that serL15A20 can lower both the tonal (at 3BPF) and separation noise compared to serL15A0, despite very slight increases in sound pressure at frequencies in excess of 4000 Hz (red circle). The broadband noise reduction can be attributed to the APG's modification of the remarkably lowered $\widehat{SAPG}$ (from serL15A0's 59.7% to serL15A20's 51.6%) shown in Fig. 8 (d) as well as $IAPG$ (by approximately 13%), while the coefficient $\beta$ is slightly reduced too. Consequently, as seen from Fig. 4, serL15A20 shows the best acoustic performance.

However, highly inclined serrations ($\geq 40°$) can be problematic for both tonal and broadband noise. Catastrophic increases in sound pressure occur at frequencies over 4k Hz for serL15A40, while these can occur even from 2k Hz for serL15A60 as observed in the red circle in Fig. 8 (c). Meanwhile, the noise reduction capability of the serrations at middle frequencies (100–3k Hz) is harmed as well, which is particularly obvious for serL15A60. As for the tonal noise, the noise rises noticeably at BPF with the increase in inclination, while 40° becomes the optimum in lowering the noise at 3BPF. Consequently, as shown in Fig. 4, serL15A40 still enables an overall noise reduction of 0.8 dB, although the SNL of serL15A60 exceeds the basic design marginally. Therefore, large inclinations can cause severe side effects at middle-to-high frequencies of over 4k Hz, counteracting the acoustic benefits of the serrations. There exists an optimum that can be found in Fig. 4 or Fig. S3 (c) for the inclination angle, which is approximately 20° for overall sound levels.

$\widehat{SAPG}$ fails to reflect the separation severity for designs with heavily inclined serrations (serL15A40 and serL15A60), while it successfully evaluates the separation severity (or noise) trends for interval and length variations, consistent with Fig. S3 (a) and (b). Fig. 8(f) suggests that an increase in inclination can considerably lower $\widehat{SAPG}$, which should lead to a lower sound pressure at least within 100-2k Hz where the inclination-induced drastic increase in noise is not observed. In contrast, the broadband noise increases slightly over this range when the inclination exceeds 40° compared to serL15A20 (Fig. 8 (c)). Furthermore, the sharp increase of $\beta$ (0.88 & 1.07) over even 0.86 of the basic design level also indicate a very even APG distribution, which may point to a different sound production mechanism.

Figs.9 (a) and (b) are the computed sound spectra by FWH analogy with PBM for noise from the blade and hub, and the blade suction face, respectively. Similar conclusions can be derived from Fig.9 (a) as from Fig.8 (c) that the large inclination of serL15A40 harms the separation noise reduction efforts over 100-2k Hz, and causes considerable increase in the noise from around 2k Hz (slightly different from the 4k

Hz in Fig.8 (c)), which suggests the reliability of PBM to analyze the noise concerning inclination. Moreover, Fig.9 (b) shows quite similar trends and high coherence with Fig.9 (a) over the above two ranges, indicating that the suction face holds the key to understanding the aeroacoustic characteristics of the blade as pointed out by the literature.[12]

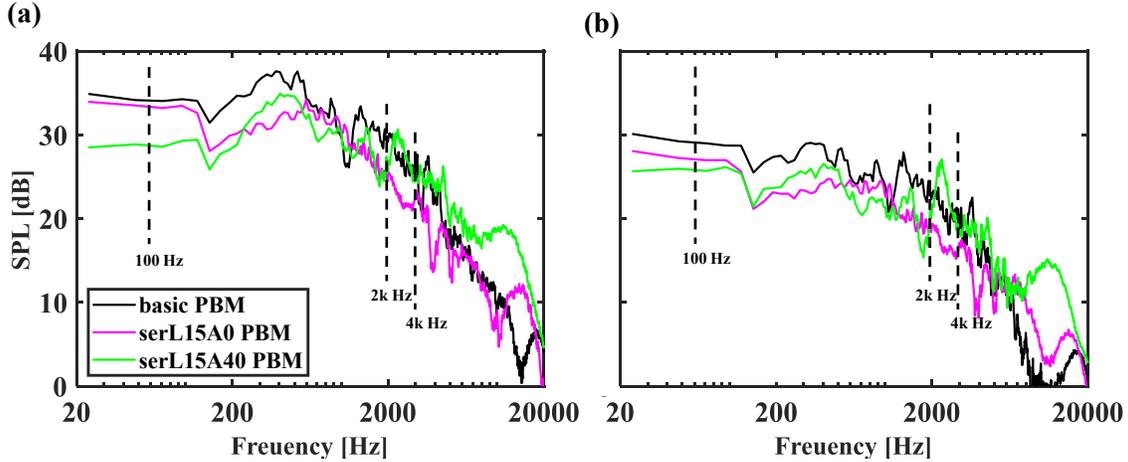

**FIG. 9.** Sound pressure level spectra of three designs via LES-based FWH analogy with PBM for (a) blade and hub, and (b) only the blade suction face.

## C. Impact of inclination of serrations

### 1. Flow-buffering effect

Figs. 10-12 are the results of the steady RANS analyses with CBM for serL15A0, serL15A40 and serL15A60, depicting the variations of velocity, pressure, and vortex field above the blade suction surfaces. The design serL15A20 shares the same trend with the increase in serration inclination and is omitted for conciseness.

A new passive flow-control mechanism is discovered that may account for the benefits of a slight inclination. As demonstrated in Figs. 10(a)-(c), the inclination forces the air seeping above the serrations to move in the $-X_1$ direction, i.e., the original momentum of oncoming flow (in the $-X_2$ direction) is partially transferred to the $-X_1$ direction. Meanwhile, the inclination also increases the confronting area of the serrations ($X_2$ direction) with the fluid, promoting the consumption of the kinetic energy of the oncoming flows. Consequently, a decelerated chordwise flow region forms above the inclined serrations (circles in Figs. 10(d)-(f)) and downstream of LE (rectangles in Figs. 10(d)-(f)), which we herein term as the 'flow-buffering' effect. It is known that slotted serrations can induce high local incidence angles at LE,[12] substantially mitigating the stagnation, while this flow-buffering effect further eases stagnation by reducing the momentum of the oncoming flow, which effectively modifies the APG downstream (see the rectangles in Figs. 11 (a) vs. (b) or (c)). Therefore, the inclination of the LE serrations further alleviate the stagnation at LE by buffering the oncoming chordwise flow, which accounts for the better aeroacoustic performance of serL15A20 than that of serL15A0 (Fig. 4 and Fig. S3(c)).

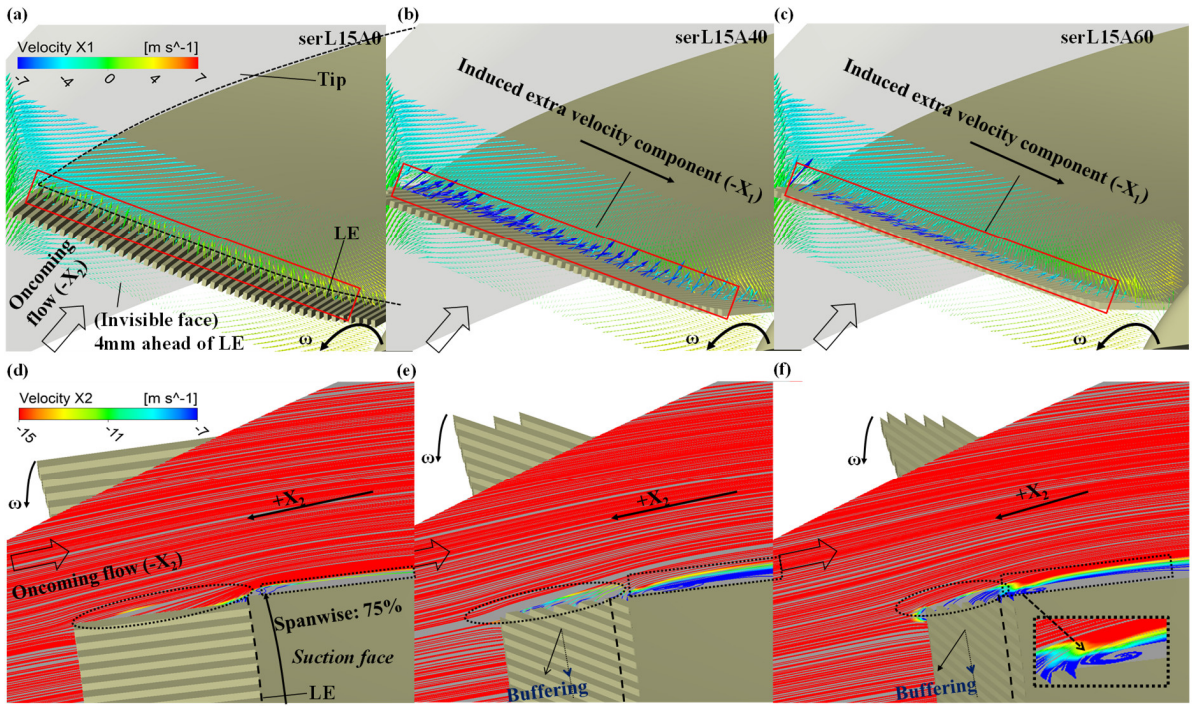

**FIG. 10.** Relative velocity vectors (in rotational coordinate frame) on a surface which is vertical to serrations and 4 mm ahead of LE (upper row) for (a) serL15A0, (b) serL15A40 and (c) serL15A60; 2D streamline (of relative velocity) on a chordwise surface at 75% blade-span (lower row), for (d) serL15A0, (e) serL15A40 and (f) serL15A60.

Nevertheless, over-buffering can be induced due to the large inclination, leading to the formation of an additional tip vortex. The flows rolling up through blade-shroud clearance from the pressure side of the blades can form a tip vortex. The inevitable $PG_1$ pairs are present in the red dashed circles of Figs. 12 (a)-(c), which drives the formation of the common tip vortex regardless of the serrations (red dashed circles in Figs. 12 (d)-(f)), serving as an important noise source for this fan (Figs. 6 (a) & (b)). However, an additional PG$_1$ pair can also be generated at the blade tip by the highly inclined LE serrations, as seen in the red solid circles of Figs. 12(b) and (c), while this is hardly observed for serL15A0 (Fig. 12(a)). This additional PG$_1$ pair thus drives the formation of the additional tip vortex (solid circles in Figs. 12(d) vs. Fig. 12(e) & (f)).

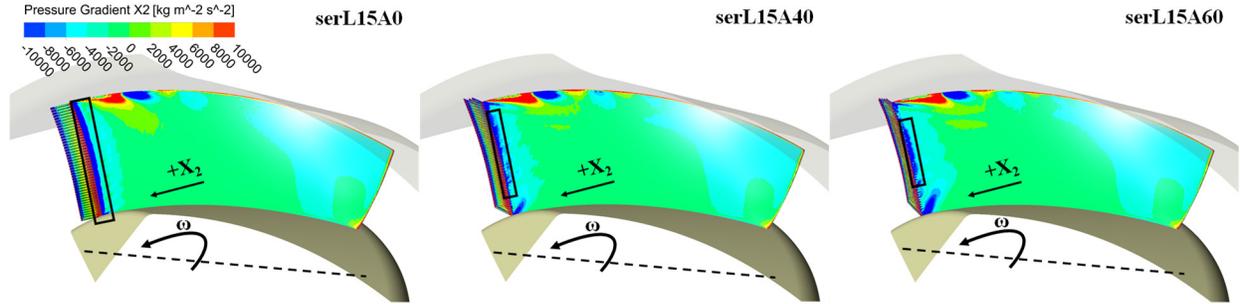

**FIG. 11.** Pressure gradient on suction faces in the chordwise direction (X$_2$) for (a) serL15A0, (b) serL15A40 and (c) serL15A60.

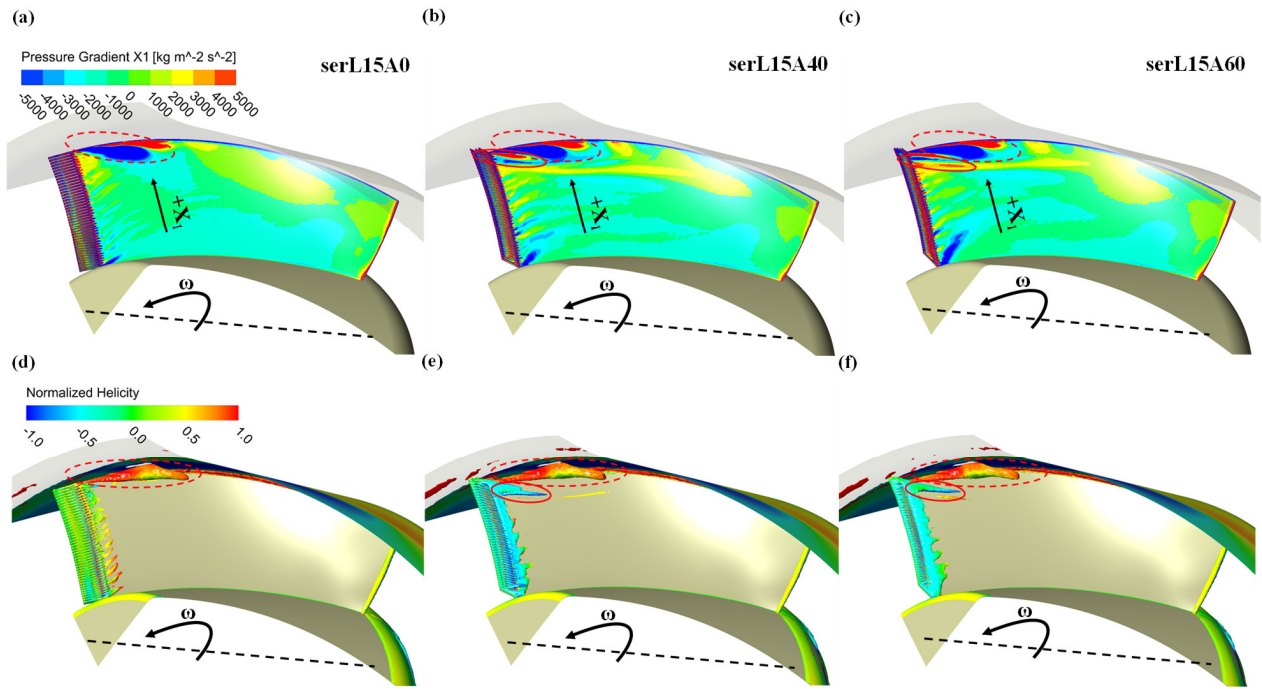

**FIG. 12.** Pressure gradient on suction faces in the spanwise direction (X$_1$) for (a) serL15A0, (b) serL15A40 and (c) serL15A60; iso-vorticity surfaces (via steady analyses, swirling strength = 1200 s$^{-1}$) in terms of normalized helicity for (d) serL15A0, (e) serL15A40, and (f) serL15A60.

These additional $PG_1$ pairs are already developed upstream of the LE above the serrations (Fig. S4 (a) vs. Fig. S4 (b) & (c), supplementary material). The formation of the additional tip vortex may be explained by the fact that the flows below heavily-inclined serrations are obstructed in the $X_3$ direction (Fig.3 (c), vertical to LE region) due to the consumption of kinetic energy (circles in Figs. 10(d)-(f)), whereby the serrations act as an extended plate above which an additional vortical structure can develop. As we know, the formation of the tip vortex can cause high frequency noise,[9, 10] which may account for the drastic increase in noise at middle-to-high frequencies (> 4k Hz) and will be discussed in section IV.C.2.

In addition, such over-buffering could worsen the chordwise flow stability as well when the inclination angle is extremely large, e.g., 60°. As seen in Fig. 10(f), a separation bubble forms right after the LE (enlarged portion), and the APG in midspan shows a slight increase as shown in the black rectangles on Figs. 11(b) vs. (c). As the LES-based FWH analogy for serL15A60 is not carried out in this study, it is not clear how this instability can affect the SNS of the blade surface.

**2. High-coherence turbulence impingement noise**

LES analyses along with the FWH analogy are employed to explore correlations between the near-field flow structures and surface noise sources under serrations with heavy serration inclination (≥ 40°). Figs. 13 (a)-(f) (Multimedia views) are the transient vortex fields and SNS for the basic design, serL15A0, and serL15A40. A lag of 3.056e-3s is used for SNS behind the corresponding vortex field of each design for better visual effect considering sound wave propagation from the blade tip of LE to the receiver. Hence, Figs. 13 (d)-(f) show the approximate synchronous sound radiation conditions under vortex fields of Fig. 13 (a)-(c), although different points on the blade surface do not share exactly identical propagation times.

We found that the sound radiating from blades depends largely on the vortex shedding patterns. Fig 13. (a) and (b) (Multimedia views) show the KH instability suppression with serrations where the strong vortical structures are broken up into small eddies downstream of LE.[12, 32] In addition, Fig 13. (b) and (c) clearly demonstrate that the flow-buffering effect further reduces the eddies in the LE region (rectangles). However, the additional tip vortex shedding due to heavy serration inclination causes strong turbulent flows to impinge on the outer surface of the blade (above the dashed line in Fig 13. (c)) as well. On the other hand, the sound emissions from the suction surface are considerably consistent with the flow field phenomena. It can be observed that transient SNS declines significantly with LE serrations (Fig 13. (d) vs. (e) or (f)). Moreover, the serration inclination results in further SNS reduction as well, especially in the LE region (rectangles in Fig 13. (e) vs. (f)) and the outer surface of the blade shows high SNS (Fig 13. (f)). To sum up, the transient SNS is largely dependent on the intensity and pattern of impingement of the turbulent flows, and LE-serrated blades lower noise mainly by suppressing turbulent flows in the mixed flow fan.

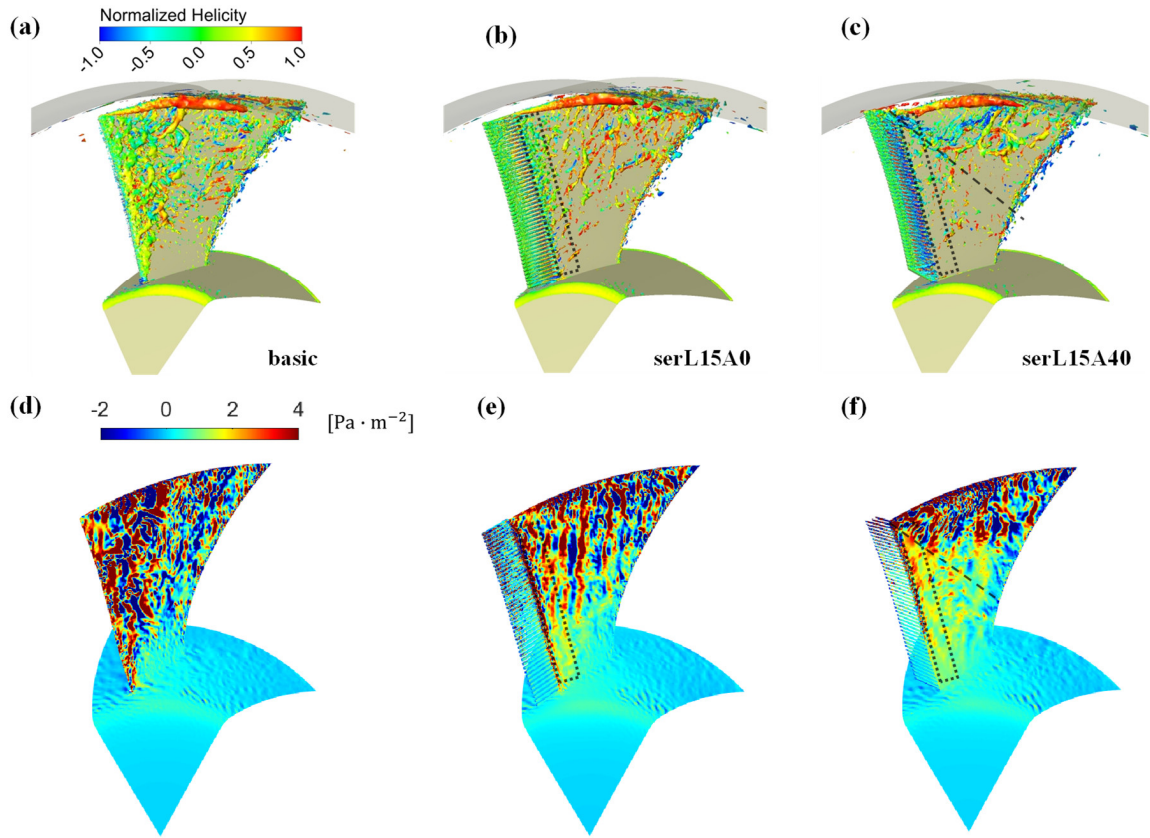

**FIG. 13.** Iso-vorticity surfaces (instant of 3.17T or 0.15652s, swirling strength = 3000 s$^{-1}$) in terms of normalized helicity for (a) basic design, (b) serL15A0 and (c) serL15A40; surface noise strength (instant of 0.15958s) and for (d) basic design, (e) serL15A0 and (f) serL15A40. (Multimedia views)

To clarify the heavy inclination's aeroacoustic effects over different frequency ranges, we conducted the discrete Fourier transform (DFT) for *SNS* (over 3.17T-4.17T) for each point (by spatial interpolation) on the blade surface and computed their frequency-averaged values over two frequency ranges by referring to Fig. 8(c) and Fig. 9, i.e., 100 – 2k (the middle frequencies, 2k-3k is skipped) and 4k – 20k Hz (middle-to-high frequencies) as shown in Fig. 14. We found that it is still the additional tip vortex induced by heavy inclination that affects the sound production over both ranges. As shown in Figs. 14 (c) and (f), the severe SNS mainly exists above the dashed line or the blade outside part, where turbulent flows (from the additional tip vortex bursting) mainly impinge. This SNS distribution pattern is entirely different from those in the remaining figures.

However, SNS in the frequency domain shows conflicting results with SPL spectra or one-third octave analyses at frequencies of over 4k between the basic design and serL14A40, which is suspected related to the coherence effect. As seen from the Fig. 14 (d) vs. (f), serL15A40 owns less powerful noise source over 4k-20k Hz than the basic design, disagreeing with the drastic noise increases within this range observed in Fig. 8 (c) or Fig. 9. This conflict should be attributed to the high coherence originating from the additional tip vortex shedding. As pointed out by a recent research, the far-field noise of an airfoil at a single frequency is the sum of uncorrelated sources that come from similar flow structures of certain coherence length passing over the trailing edge.[14] The additional tip vortex of serL15A40 easily bursts into turbulent flows of similar sizes which also easily impinge onto the same positions on the blade surface, scattering into the sound, whereas noise of the basic design is produced by turbulence impacts separated from the entire LE whose sizes vary spanwise due to different rotational linear velocity from the blade root to the tip. Therefore, the surface noise sources in the frequency domain should also be examined in terms of coherence as well aside from SNS.

We found that the high coherence of the turbulence impingement noise should mainly account for the noise level increases at frequencies of over 4k Hz caused by heavy serration inclination (over 40°). Fig. 15 depicts each design's frequency-averaged (magnitude squared) coherence over the same frequency ranges as in Fig. 14, which is conducted between the far-field sound pressure ($p_f$) and the local contributions ($dp_f/dS$) from blade suction surface. In Fig. 15, the coherence over 4k Hz considerably rises due to serrations appending (Fig. 15 (d) vs. (e)), which further increases significantly accompanying heavy inclination angle (Fig. 15 (e) vs. (f)). Hence, it is very likely that the strong local coherence of serL15A40 (circle in Fig. 15 (f)) enlarges the acoustic impact of corresponding surface noise sources (circle in Fig. 14 (f)), accounting for the aforementioned conflict.

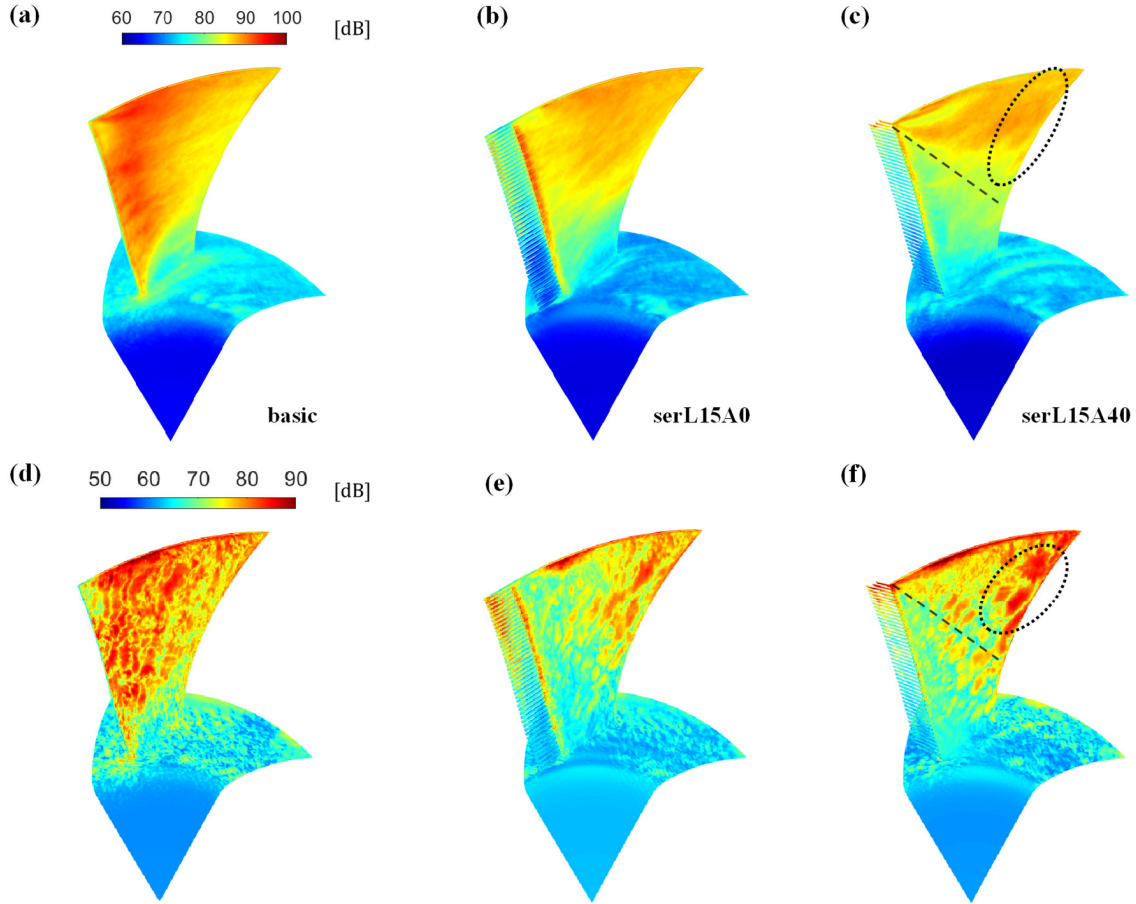

**FIG. 14.** Frequency-averaged surface noise strength over middle frequency range (100 – 2k, upper row) and middle-to-high frequency range (4k – 20k, lower row) for (a) & (d) basic design, (b) & (e) serL15A0 and (c) & (f) serL15A40, respectively.

Likewise, the high coherence induced by the additional tip vortex may also deteriorate noise lowering benefits of serrations over the middle frequency range. Despite obvious SNS reduction over this range (Figs. 14 (b) vs. (c)) accompanying heavy serration inclination, high coherence perhaps enlarges sound producing impact for the local surface noise sources (circles in Fig. 14 (c) and Figs. 15 (c)), accounting for that serL15A40 surpass serL15A0 at some frequencies within 100-3k Hz (Fig. 8(c) or Fig. 9). Such side effect should be more severe for serL15A60.

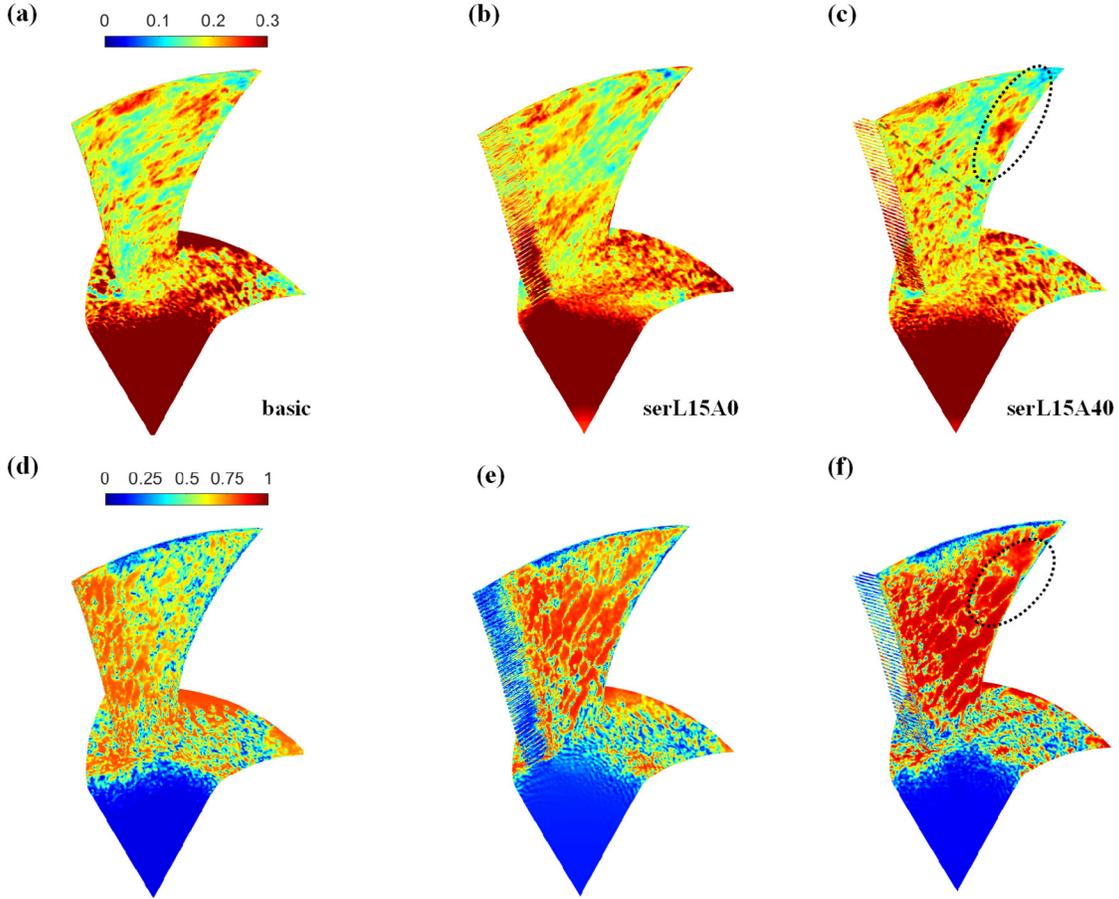

**FIG. 15.** Frequency-averaged magnitude squared coherence between $p_f$ and $dp_f/dS$ over middle frequency range (100 – 2k, upper row) and middle-to-high frequency range (4k – 20k, lower row) for (a) & (d) basic design, (b) & (e) serL15A0 and (c) & (f) serL15A40, respectively.

## D. Limitations of proposed methods

As discussed above, the SNS visualization method (proposed in section II.C) and the $SAPG$ (proposed in section IV.A) both manifest some limitations when dealing with the acoustic predictions for serL15A40 and serL15A60 with heavy inclination angles. Although the SNS visualization can accurately uncover the spatial distribution of acoustic sources on the body surface, it cannot involve effects of their coherence, which may mislead into underestimating the sound level over some frequency ranges. The $SAPG$ merely considers the effect of chordwise APG for separation evaluation without involving spanwise pressure gradient (PG$_1$) that induces the additional vortical structure, and hence the separation noise evaluation may get underestimated as well.

## V. CONCLUSIONS

In this study, we investigated the morphological effects of slotted LE serrations for a mixed flow fan in terms of aeroacoustic signatures by varying three design parameters. We also proposed two novel CFD-informed methods to visualize the surface noise strength and evaluate the separation severity. The conclusion obtained are listed below.

1) We found that the noise emitted by the basic blade is mainly from the leading-edge region as well as the blade tip, and the separation-induced LE-noise concerning flow separation can be suppressed remarkably using the LE serration. In addition, the LE serrations provide a robust tradeoff between the noise reduction and aerodynamic performance sustaining under various serration morphologies, among which only two designs show a loss in efficiency or increase in noise marginally. We also confirmed that LE serrations offer more benefits than the remaining optimization space of the prototype.

2) Reducing the interval and extending the length of the serrations can help lower the broadband noise over a broad frequency range (100-20k Hz). However, the latter is more effective for separation noise over 100–3k Hz, while a slight inclination can facilitate further reduction of this component as well. A smaller interval leads to lower tonal noise besides the broadband noise reduction, which makes narrower serrations more desirable. By contrast, length and inclination have optimal values of 15 mm and 20° for the overall noise level reduction, as larger serration lengths cause more tonal noise, while heavy inclinations ($\geq 40°$) induces broadband noise that is especially unfavorable at middle-to-high frequencies (4k -20k Hz).

3) The benefits of noise reduction brought about by the slight inclination ($\leq 20°$) can be ascribed to a passive flow control mechanism termed as the flow-buffering effect, which can also alleviate the stagnation at the LE. However, over-buffering under heavy inclinations ($\geq 40°$) results in an additional tip vortex structure and even potential flow chordwise instability ($\geq 60°$). This tip vortex can cause serious noise issues including high-coherence turbulence impingement noise, despite further SNS reduction with flow-buffering. Consequently, this counteracts the LE-serrations' aeroacoustic benefits over a broad range (100-20k Hz) and deteriorates acoustic performance of a fan, particularly at middle-to-high frequencies (4k-20k Hz).

4) The SNS visualization can help accurately determine the spatial distributions of the surface noise sources in temporal and frequency domains, but overlooks the influence of their coherence, while the index *SAPG* successfully predicts the severity of the chordwise separation with consistent APG modifications under varying serration morphologies, although it underestimates the overall severity in cases of heavy inclination due to lack of involvement of the spanwise PG.

Our work deepens the existing understanding of the morphological effects of serrations on the mixed flow fan noise. Furthermore, we hope that the methods proposed in the study will help researchers identify and evaluate noise sources more conveniently to ease the design process in the future.

## SUPPLEMENTARY MATERIAL

See the supplementary material for the verification of grid independence study on pressure gradient (Fig. S1); the verification of setup insensitivity with moving-averaged torque (Fig. S2); sound pressure level of different designs (Fig. S3); and the spanwise pressure gradient on a surface vertical to serrations and 4mm ahead of LE (Fig. S4).

## ACKNOWLEDGMENTS

This research was supported partly by *Japan Society for the Promotion of Science (JSPS)* KAKENHI No.24120007 and by an endowed project by *TERAL Inc., Japan*. JW acknowledges Excellent International Student Scholarship provided by Chiba University. JW thanks Jiaxin Rong and Jianwei Sun for their valuable help and suggestions.

## DATA AVAILABILITY

The data that support the findings of this study are available from the corresponding author upon reasonable request.

## APPENDIX

**1. Equation simplification with a node noise source**

The form of FWH equation that F. Farassat derived out (equation (9) in Ref. 51 or Eq.8) for a receiver (at $\vec{x}$) can be written as

$$p_f(\vec{x},t) = \int_{S^*} \frac{1}{4\pi} \left[ \frac{1}{c} \frac{\partial}{\partial t} \left( \frac{\rho c v_n + p_s \cos\theta}{r|1 - M_r|} \right)_{\tau^*} + \left( \frac{p_s \cos\theta}{r^2|1 - M_r|} \right)_{\tau^*} \right] dS \qquad (A1)$$

We can verify our in-house code with a noise source of revolving node at constant revolving speed (cycle: $T$) by comparing the computed $p_f$ to the analytical solution, since the last step of solving Eq. A1 is to conduct integration by simply summing up the node-based values of integrand times corresponding tethered area on the body surface. The integrand values and the tethered areas can be obtained from the discretized data of the CFD result. Therefore, the code correctness in this study relies on whether it can produce correct solution for each revolving node on blade surface. We assume a node revolving constantly is tethered with area $\Delta S$ and normal vector $\vec{n}$, and build a cartesian coordinate system, taking the revolving axis as $y_3$ and the node revolving on the plane of $y_1 O y_2$ (O is the origin), as shown in Fig. A1 (a).

For a fixed receiver locating at noise source's revolving axis, the $v_r$ equals 0 (and hence, $M_r = v_r/c = 0$) due to no velocity component in the $\vec{r}$ direction, while $v_n$ is a constant as the constant revolving velocity forms an unvaried angle with $\vec{n}$. Besides, $r$ and $\theta$ (angle between $\vec{r}$ and $\vec{n}$) are constant as well under this special condition. Thus, we can simplify Eq. A1 by substituting these values into it and get,

$$p_f(t) = \int_{S^*} \frac{1}{4\pi} \left[ \frac{\cos\theta}{cr} \frac{\partial}{\partial t} (p_s)_{\tau^*} + \frac{\cos\theta}{r^2} (p_s)_{\tau^*} \right] dS \qquad (A2)$$

As we merely study a node of tethered area $\Delta S$, the integration result is given by,

$$p_f(t) = \frac{\Delta S \cos\theta}{4\pi} [\frac{1}{cr}\frac{\partial p_s}{\partial t} + \frac{1}{r^2}p_s]_{\tau^*} \tag{A3}$$

Now we assume node noise source and the receiver are $\varphi r$ and $\psi r$ distant to the origin ($\varphi^2 + \psi^2 = 1$), respectively (Fig. A1 (a)). $\cos\theta$ can be derived if we assume $\vec{n} = [n_1\ n_2\ n_3]$ at the instant when node passes $y_1$ as in Fig. A1 (a), such as,

$$\cos\theta = \frac{\vec{n}\cdot\vec{r}}{|\vec{n}||\vec{r}|} = -\varphi n_1 + \psi n_3 = -\varphi n_1 + \sqrt{(1-\varphi^2)}\, n_3 \tag{A4}$$

As the $\theta$ is a constant, $\cos\theta$ always equals to $-\varphi n_1 + \sqrt{(1-\varphi^2)}\, n_3$. Substitute Eq. A4 into Eq. A3, we can derive out $p_f$, such as,

$$p_f(t) = \frac{\Delta S(-\varphi n_1 + \sqrt{(1-\varphi^2)}\, n_3)}{4\pi} [\frac{1}{cr}\frac{\partial p_s}{\partial t} + \frac{1}{r^2}p_s]_{\tau^*} \tag{A5}$$

where the $\tau^*$ (allowed to be negative) is sound emission time and has the relationship with sound receiving time is given by,

$$\tau^* = t - \frac{r}{c} \tag{A6}$$

## 2. Verification with equation specialization

Now we can impose a specific $p_s(\tau^*)$ on the node. A function of sinusoidal form is adopted for simplicity defined as,

$$p_s = A \cdot \sin\left(\frac{2\pi}{T}\tau^*\right) \tag{A7}$$

The $p_s(\tau^*)$ can be any function as long as it is periodic and shares an identical cycle ($T$) with the revolving, because in our study, the surface pressure fluctuation of the blade can be seen periodic accompanying the impeller rotating. Substitute Eq. A6 and Eq. A7 into Eq. A5, $p_f(t)$ can be written as,

$$p_f(t) = \{\frac{\cos\frac{2\pi}{T}(t-\frac{r}{c})}{2Tcr} + \frac{\sin[\frac{2\pi}{T}(t-\frac{r}{c})]}{4\pi r^2}\} \cdot A \cdot \Delta S(-\varphi n_1 + \sqrt{(1-\varphi^2)}\, n_3) \tag{A8}$$

Eq. A8 is the analytical formular we use to verify the code. Now we need to provide specific values for the variables on the right-hand side of Eq. A8. Thus, we take $r$=200 m, $T$=2 s, $A$=50 Pa, $\Delta S$=0.1 m², $n_1$=0.2, $n_3$=0.1, $\varphi = 0.3$ and they can be any value as long as $n_1^2 + n_3^2 \leq 1$ while $0 \leq \varphi \leq 1$.

For numerical computation with our code, it is easy to find solution of $p_f(t)$ in this case is unrelated to the initial position of the node and thus, the instant in Fig. A1 (a) is used for initialization. Time-varying variable functions on the right-hand side of Eq.8 can be obtained and derived with the Eq. A7 and the specific variable values above while $\rho$ is set to be 1.165 kg/m³. 1300 is taken as the time step number to calculate discretized data as the inputs into the code. It needs to be noted that our code can solve far-field noise radiating from randomly moving surface and thus, conditions of $M_r = 0$ and constant $v_n$ are not incorporated, although verification herein is limited to cases of surface revolving at a constant speed.

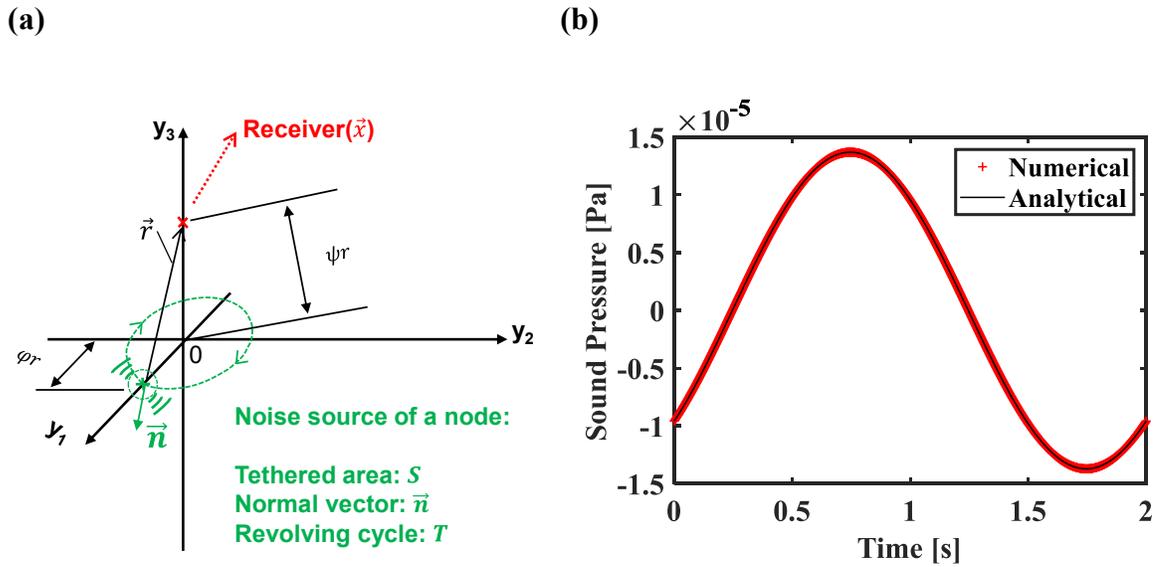

**FIG. A1.** (a) Illustration for a sound source of revolving node and a receiver locating at the revolving axis, and (b) sound pressure vs. time through numerical and analytical computations (time step number is 1300).

Fig. A1 (b) shows the far-field sound pressure results within 1 T of the numerical solution by the code and the analytical solution to Eq. A8 and they highly overlap with each other. The maximum numerical error is below 1.5% of the analytical result. Fig. A2 is the sensitivity test of the time step number for the maximum numerical error which is found below 0.7% exceeding 1500 time steps. The time step number we recorded in LES analyses for our blades are 1677 and 2097 for the CBM of the basic design and other three PBMs, respectively. Thus, the correctness and accuracy of our results are ensured.

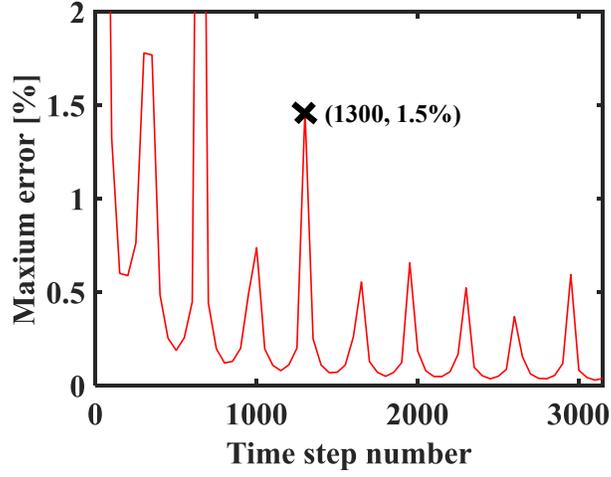

**FIG. A2.** Maximum numerical error vs. adopted time step number.

### 3. Classification of the noise source in our study

It needs to be noted that the Eq. 8 can be divided into two parts according to Farassat's work given by,

$$p_{f_1}(\vec{x},t) = \frac{1}{4\pi}\int_{S^*}[\frac{1}{c}\frac{\partial}{\partial t}\left(\frac{\rho c v_n}{r|1-M_r|}\right)_{\tau^*}]dS \tag{A9}$$

$$p_{f_2}(\vec{x},t) = \frac{1}{4\pi}\int_{S^*}[\frac{1}{c}\frac{\partial}{\partial t}\left(\frac{p_s cos\theta}{r|1-M_r|}\right)_{\tau^*} + \left(\frac{p_s cos\theta}{r^2|1-M_r|}\right)_{\tau^*}]dS \tag{A10}$$

where $p_{f_1}$ is the thickness noise or the monopole source that only requires knowledge of kinematics and geometry of the body surface, while $p_{f_2}$ is the loading noise or the dipole source that is related to the pressure fluctuation on the surface. Obviously, $p_{f_1} = 0$ when the receiver is at the revolving axis as $v_n$ is constant. Therefore, in our study for the mixed flow fan, the noise we computed at the corresponding location in the experiment (Fig.2(b)) contains only the dipole component.